\documentclass[prc,showpacs,preprintnumbers,amsmath,amssymb]{revtex4}
\usepackage{graphicx}% Include figure files
\usepackage{dcolumn}% Align table columns on decimal point
\usepackage{amsmath}% bold math
\renewcommand{\vec}[1]{\boldsymbol{#1}}

\begin{document} 
\title{\Large 
Towards a practical approach for  self-consistent
large amplitude collective motion\footnote{Dedicated to the memory of 
Professor Abraham Klein}} 
\author{Daniel Almehed}
\email{D.Almehed@umist.ac.uk}
\author{Niels R.~Walet} 
\email{Niels.Walet@umist.ac.uk}
\affiliation{Department of Physics, UMIST, P.O. Box 88, 
Manchester M60 1QD, United Kingdom} 
 
\begin{abstract}
We investigate the use of an operatorial basis in a self-consistent
theory of large amplitude collective motion. For the example of the
pairing-plus-quadrupole model, which has been studied previously at
equilibrium, we show that a small set of carefully chosen 
state-dependent basis operators is sufficient to approximate the exact
solution of the problem accurately. This approximation is used to
study the interplay of quadrupole and pairing degrees of freedom along
the collective path for realistic examples. We show how this
leads to a viable calculational scheme for studying nuclear structure,
and discuss the surprising role of pairing collapse.
\end{abstract} 
\maketitle 

\section{Introduction}
There is a general quest for understanding of complicated phenomena in
terms of a limited set of degrees of freedom, chosen through some
method appropriate for the problem at hand. Many approaches are
available, in areas ranging from field theories to atomic physics
(see, e.g., the reviews in Ref.~\cite{To98}). These are typically
based on the concept of ``relevant degrees of freedom'', or on the
introduction of collective motion and collective paths -- which are
two ways to express rather similar principles!

More specific to the nuclear problem studied in this paper, the old
question ``what is the correct choice of collective coordinate in
a many-body system'' has had quite a few partial answers, see the
review~\cite{DK00} for a discussion of some of these. The holy grail
of this approach is a method that determines a collective path
self-consistently, based only on knowledge of the Hamiltonian
governing the system. Preferably the method chosen should allow us to
measure whether the limited dynamics in a few coordinates makes sense
at all, or in the language used above, address the question ``how
effective are the effective degrees of freedom?''.

The constrained Hartree-Fock-Bogoliubov method is commonly used to
describe collective paths in nuclear physics (see
e.g.~\cite{BK67,RS80}).  This approach, where the collective subspace
is generated by a small number of one-body constraints, also goes by
the name of generalised cranking. The one-body constraints usually
consist of a few carefully chosen multipole (particle-hole) operators,
as well as a few generalised pairing (particle-particle) ones.  For
large scale realistic problems such as the description of nuclear
fission the number of generalised cranking operators needed in order
to make a realistic calculation becomes very large.  There is also no
reason to limit the constraints to the standard choices; other degrees
of freedom, especially those involving spin-orbit interactions might
also be important. A more satisfactory method should allow the cranking
operators to be determined by the nuclear collective dynamics itself.

One such approach, followed in this paper and set out in detail in the
review paper~\cite{DK00} (a similar approach, plus relevant references, 
can be found in Ref.~\cite{KN03}), 
leads to a very well-defined approach, which
can in principle be solved knowing the Hamiltonian and model space.
To find the adiabatic collective path we use the local harmonic
approximation (LHA).  It consists of a constrained mean-field problem
that needs to be solved together with a local random phase
approximation (RPA), which determines the constraining operator. This
approach lacks practicality, since the size of the RPA problem is, for
a system with pairing, proportional to the size of the single-particle
space squared. Even though enormous matrices can routinely be
diagonalised on modern computer systems, the algorithm requires
repeated diagonalisation of such a matrix, which makes an
implementation in realistic calculations prohibitively time consuming.

This requires a solution, or at least a good approximation, and this
is the subject of the present paper. A first approach to solving the
problem has been suggested in the work of one of the
authors~\cite{NW99}.  The best way to test such ideas is to use a
semi-realistic model, where approximations can be tested against the
full method, such as the pairing-plus-quadrupole model as employed by
Barranger and Kumar~\cite{BK68} in their seminal work. It has been
shown in the past~\cite{NW99} that at equilibrium the RPA can be
solved quite efficiently using a simple basis of operators. Related
work by Nesterenko \emph{et al.}~\cite{NK02} may also have some
bearing on this problem, but will not be investigated here. In short,
the idea is that the basic operators of the model, weighted by a
suitable power of the quasi-particle energies, give excellent
results. The state dependence induced by the quasi-particle energies
is crucial to the success of the approach, and is the main difference
with methods based on ``naive'' constraints. Since the original work
was only done at equilibrium, we must still check that such a basis of
operators provides a good solution along the collective path, and we
indeed find some important modifications to the method discussed
in~\cite{NW99}. Once a collective path has been found we can
diagonalise the collective one-body Hamiltonian along this path,
including all the zero modes arising from broken symmetries. This will
give information on how the collective motion influences the ground
state properties of the nuclei.  In several of the examples discussed
below we find low lying states with pairing collapse which influence
the collective behaviour of the system. If we now quantise the
collective dynamics, we \emph{must} include the pairing
rotations. This is due to the fact that a point with collapsed pairing
behaves similarly to the origin in polar coordinates, with the pairing
phase playing the role of the polar angle.  In this work we shall only
study the effect of the pairing-rotational modes, ignoring collective
rotation for the time being.

The paper is organised as follows. In Sec.~\ref{sec:Formalism} we briefly
present the basic principles of our approach, to highlight those issues
that will make the results easier to understand. The practical form
of the equations for the type of many-body problem considered here is
also discussed, and the form of the approximation is introduced.
Results are then given in Sec.~\ref{sec:Results} and finally we draw
some conclusions in Sec.~\ref{sec:Conclusions}.

\section{Formalism\label{sec:Formalism}}

The formalism, as set out in detail in~\cite{DK00}, is based on
time-dependent mean field theory, and the fact that a classical
dynamics can be  associated with it. The issue of selecting
collective coordinates, and determining their coupling to other
degrees of freedom, thus becomes an exercise in classical mechanics.
Furthermore, if we assume adiabaticity, a slow motion where the
Hamiltonian can be expanded to second order in momenta, we have a
problem that can be solved. The solution can be stated without any direct
reference to the original nuclear many-body problem and the choice of 
the interaction.

\subsection{Local harmonic approximation for the collective path in the 
	adiabatic limit\label{sec:LHA}}

We assume a classical Hamiltonian depending on a set of real canonical
coordinates, $\xi^\alpha$ ($\alpha=1,\ldots ,N$), and conjugate
momenta, $\pi_\beta$ ($\beta=1,\ldots ,N$), of the form ($\xi$ and
$\pi$ thus parametrise $\left| \Psi \right>$ \cite{DK00})
\begin{equation}
	\label{eq:H1}
	{\cal H}(\xi,\pi) = \left< \Psi \right| H \left| \Psi \right> .
\end{equation}
We shall use a tensor notation, where we use upper indices for coordinates and lower 
indices for momenta. When the same symbol appears as both upper and lower index there 
is an implicit sum over that index.

The potential $V(\xi)={\cal H}(\pi=0)$ and the mass matrix $B^{\alpha \beta}$
are given by an expansion of ${\cal H}(\xi,\pi)$ in powers of $\pi$ in zeroth
and second order, respectively,
\begin{equation}
	\label{eq:H2}
	{\cal H}(\xi,\pi) = V(\xi) + \frac{1}{2} B^{\alpha \beta} \pi_\alpha 
	\pi_\beta + {\cal O}(\pi^4) .
\end{equation}
Terms of higher order (such as $\pi^4$) are supposed to be negligible.
The kinetic energy in the Lagrangian formalism contains the inverse
$B_{\alpha\beta}$ of the mass matrix, 
$K=\frac{1}{2} \dot \xi^\alpha B_{\alpha\beta} \dot \xi^\beta$, and
can be interpreted as an inner-product in the tangent space to a curved
manifold.  The inverse of the mass matrix $B^{\alpha \beta}$, is thus the 
metric tensor; in other
words the matrix $B_{\alpha\beta}$ represents the Riemanian geometry in
configuration space, since it measures lengths in the tangent
space. This clearly would not be the case if we had higher order terms
in the kinetic energy.

The central part in our approach to large amplitude motion is a search
for collective (and non-collective) coordinates $q^\mu$ which are
obtained by an invertible point transformation of the original
coordinates $\xi^\alpha$, preserving the quadratic truncation of the
momentum dependence of the Hamiltonian~\footnote{In Ref.~\cite{DK00}
it is discussed how such a ``truncation before transformation'' must
be replaced by a ``truncation after transformation'' for certain
exactly solvable multipole models. Complication arising from such an
approach are such that we shall ignore such an extension here.},  by
\begin{equation}
	\label{eq:fg1}
	q^\mu = f^\mu(\xi), \quad \xi^\alpha = g^\alpha(q) \quad \left( 
	\mu , \alpha =  1,\ldots ,n \right) ,
\end{equation}
and the corresponding transformation relations for the momenta
$p_\mu$ and $\pi_\alpha$,
\begin{equation}
	\label{eq:fg2}
	p_\mu = g^\alpha_{,\mu}\pi_\alpha, \quad \pi_\alpha = 
	f^\mu_{,\alpha} p_\mu 
\end{equation}
where we use a standard notation for the derivatives,
$g^\alpha_{,\mu} \equiv \frac{\partial}{\partial q^\mu} g^\alpha$ and 
$f^\mu_{,\alpha} \equiv \frac{\partial}{\partial \xi^\alpha} f^\mu$.
The adiabatic Hamiltonian, Eq.~(\ref{eq:H2}), is then 
transformed into
\begin{equation}
	\label{eq:H3}
	\bar{{\cal H}}(q,p) = \bar{V}(q) + \frac{1}{2} \bar{B}^{\mu \nu} 
	p_\mu p_\nu + {\cal O}(p^4) 
\end{equation}
in the new coordinates. The new coordinates $q^\mu$ are now to be divided
into three  categories: the collective coordinate $q^1$, the
zero-mode coordinates $q^I$, $I=2,\ldots,M+1$, which describe motions that
do not change the energies and finally the non-collective
coordinates $q^a$, $a=M+2, \ldots ,n$. [The approach can easily be
generalised to include more than one collective coordinate, but that
will not be discussed here.]

The collective coordinate is determined by means of the
solution to the local harmonic approach, 
which consists of a set of self-consistent
equations.  These are:
\begin{enumerate}
\item The force equations
\begin{equation}
	\label{eq:force1}
	\bar{{\cal H}}_{,\alpha} = \Lambda f_{,\alpha} + \Lambda_I f^I_{,\alpha} ,
\end{equation}
where $f^I$ are the zero-modes (also called Nambu-Goldstone or
spurious modes) and $\Lambda_I$ represents a set of Lagrange
multipliers (which in nuclear physics are usually called cranking
parameters).  $\Lambda$ is a Lagrange multiplier for the collective
mode, stabilising the system away from equilibrium (we shall often
denote it as the generalised cranking parameter).\
\item The local RPA equation
\begin{equation}
	\label{eq:localRPA1}
	\bar{V}_{;\alpha \gamma} B^{\gamma \beta} f_{,\beta} = 
	\left( \hbar \Omega \right)^2 f_{,\alpha} ,
\end{equation}
where the covariant derivative $V_{;\alpha \beta}$ is defined in the
usual way ($V_{,\alpha \beta}=(V_{,\alpha})_{, \beta}$),
\begin{eqnarray}
	\label{eq:covar}
	V_{;\alpha \beta} &\equiv& V_{,\alpha \beta} - \Gamma^\gamma_{\alpha \beta}
	V_{,\gamma} , \\
	\label{eq:Cristoffel}
	\Gamma^\alpha_{\beta \gamma} &=& B^{\gamma \delta} \left( 
	B_{\delta \beta ,\gamma} + B_{\delta \gamma ,\beta} - 
	B^{\beta \gamma ,\delta} \right) .
\end{eqnarray}
Zero modes correspond to zero eigenvalues of the RPA. In principle
great care needs to be taken to have zero modes behave correctly away
from equilibrium. The symplectic RPA~\cite{DK00} is the correct way to
do so; unfortunately it is rather cumbersome, and as a practical
approximation we shall ignore the corrections arising from this
approach here. In this paper we will also neglect the covariant corrections
to the RPA, since they are time-consuming to calculate. This means that we
do not treat the zero-modes absolutely correctly.
\end{enumerate}
The collective path is found by solving Eqs.~(\ref{eq:force1})
and~(\ref{eq:localRPA1}) self-consistently, i.e., we look for a path
consisting of a series of points where the lowest non-spurious
eigenvector of the local RPA equations also fulfils the force
condition.  In the minimum of the potential the spurious solutions
decouples from the other collective and non-collective solutions. When
we are following the collective path away from the minimum one can use
special techniques, called the symplectic version of the theory, to
avoid mixing of the spurious solution and the physical
solutions~\cite{NW98}. This has the disadvantage that it is
numerically much more difficult to implement. In this paper we have
chosen to ignore the effects of the spurious admixtures to the RPA
wave-functions, but these are expected to be small, at least close to
the minimum. As a result there will be a finite overlap between the 
collective coordinate and the spurious operators.
The price paid for these approximations is that at points
where RPA frequencies should cross we get narrow avoided
crossings. The narrowness is a measure of the severity of the
truncation errors. One way to by-pass such problems, is to use a basis
of operators, where such crossings are extremely rare.

\subsection{Large amplitude collective motion with local harmonic approximation
\label{sec:LACM}}
The local harmonic approximation has been described in~\cite{DK00}.
There the structure is discussed in great detail, as is the transition
between nuclear physics and classical mechanics. Here the formalism
is extended to include pairing and constraints on particle number. We
start with the time-dependent Hartree-Fock-Bogoliubov equations; in
this case one finds that a natural choice for the coordinates $\xi$
and $\pi$ are the real and imaginary parts of the generalised density
matrix in its locally diagonal form~\footnote{We work in a local basis,
that changes from point to point along the collective path, where at
each point the coordinates and momenta parametrise the deviation of
the generalised density matrix from the diagonal form that specifies
the current point.}, where the change in the pairing density, 
${\cal K}_{q q'}=\left< \Phi \right| a_{q'} a_q \left| \Phi \right>$,  can be 
parametrised as~\cite{DK00}
\begin{equation}
	\label{eq:xipi}
	{\cal K}_\alpha = \frac{1}{\sqrt{2}} \left( \xi^\alpha + i 
	\pi_\alpha \right) .
\end{equation}

We want to find a solution of the local RPA equation, $f^{\text{new}}$, at the 
generalised density ${\cal R}^{\text{new}}$ satisfying the generalised cranking 
equation (\ref{eq:force1})
\begin{equation}
	{\cal H}\left[{\cal R}^{\text{new}}\right]_{q q'} - \lambda f^{\text{new}}_{q q'} - 
	\sum_{\tau=n,p} \mu N^{\text{new}}_{\tau\,q q'}= 0 ,
	\label{eq:cranking1}
\end{equation}
where $f^{\text{new}}$ is also an eigenvector of the RPA 
equation~(\ref{eq:localRPA1}) at ${\cal R}^{\text{new}}$ and $N_{n,p}$ are  the particle 
number operators for neutrons and protons, respectively.
We use $f$ at the previous point as input, and try to find a point a
fixed length $\Delta Q$ further along the path, that satisfies 
Eq.~(\ref{eq:cranking1}) for the ``old'' value of $f$. Subsequently, a new $f$ 
is found by solving the RPA equations, and this procedure is repeated until
Eqs.~(\ref{eq:localRPA1}) and~(\ref{eq:cranking1}) are satisfied simultaneously.

The cranking equation 
\begin{equation}
	{\cal H}\left[{\cal R}^{(n)}\right]_{q q'} - \lambda^{(n)} f^{(n-1)}_{q q'} - 
	\sum_{\tau=n,p}\mu N^{(n)}_{\tau\,q q'}= 0 ,
	\label{eq:cranking2}
\end{equation}
is solved with the additional constraint that
\begin{equation}
	\Delta Q = \left( f^{(n)} + f^{(0)} \right) \cdot \left( 
	{\cal R}^{(n)}_i - {\cal R}^{(0)} \right)
	\label{eq:dQ1},
\end{equation}
is fixed ($\cdot$ represent a scalar product). The initial values $f^{(0)}$ and ${\cal R}^{(0)}$ 
are the results obtained at the previous self-consistent point on the 
collective path. $\Delta Q$ is a measure of the step length in the collective coordinate 
and Eq.~(\ref{eq:dQ1}) is actually a linear approximation to the 
integral definition of the change in collective coordinate
\begin{equation}
	\Delta Q = \int^{{\cal R}_1}_{{\cal R}_0} {\rm Tr} 
	\left( f \delta {\cal R} \right) .
	\label{eq:dQ2}
\end{equation}
The value of $\Delta Q$ depends on the normalisation of $f$. In the following 
we choose the normalisation in such a way that the collective mass, $\bar B$, 
is position independent,
\begin{equation}
	\label{eq:CMass}
	\bar B=f_\alpha B^{\alpha \beta} f_\beta = 1 .
\end{equation}
Equation~(\ref{eq:cranking2}) is solved iteratively by a
constrained minimisation, where the change of the generalised density 
in the $i$th step of the mean-field iteration is given by
\begin{eqnarray}
	\label{eq:dRni1}
	\Delta_i^{(n)} {\cal R} &=& \Delta_{i-1}^{(n)} {\cal R} + \epsilon_i^{(n)}
	\frac{f^{(n)}}{f^{(n)} \cdot f^{(n)}} + \sum_{\tau=n,p} \eta^{(n)}_{i \tau}
	\frac{N^{20(n)}_{i \tau}}{N^{20(n)}_{i \tau} \cdot N^{20(n)}_{i \tau}}
	+\Delta_{\perp i}^{(n)} {\cal R} , \\
	\label{eq:dRni2}
	\Delta_{\perp i}^{(n)} {\cal R} &=& \delta^{(n)}_i \left\{ H^{20(n)}_i -
	\lambda^{(n)}_i f^{(n)} - \sum_{\tau=n,p} \mu^{(n)}_{i \tau} 
	N^{20(n)}_{i \tau} \right\} ,
\end{eqnarray}
where $\Delta^{(n)}_i {\cal R} = {\cal R}^{(n)}_i - {\cal R}^{(0)}$. 
The step length in the mean-field iteration $\delta_i^{(n)}$ is chosen to be 
small for small $i$, to make sure that the iterations converge, but 
can be chosen larger as the iteration approaches the minimum of the constrained 
mean-field. The parameters $\epsilon$, $\eta$, $\lambda$
and $\mu_\tau$ are calculated from the set of conditions discussed below.
For each $n$ the $i$-iteration is initiated by choosing 
\begin{equation}
	\Delta_0^{(n)} {\cal R} = \epsilon_0^{(n)} \frac{f^{(n)}}{f^{(n)} \cdot 
	f^{(n)}} + \sum_\tau \eta_{0 \tau}^{(n)}\frac{N^{20(n)}_{i \tau}}{
	N^{20(n)}_{i \tau} \cdot N^{20(n)}_{i \tau}} .
	\label{eq:dRn0}
\end{equation}

There are two types of constraints that give the undetermined parameters in 
the method described above: the fixed size of the steps in the collective 
coordinate~(\ref{eq:dQ1}) and the constraint on particle number.
The particle numbers are constrained by requiring that $\Delta {\cal R}$ does not 
change the expectation values $N_\tau$. Such a constraint can be written in differential form as
\begin{equation}
	\Delta^{(n)}_i {\cal R} \cdot N^{20(n)}_{i \tau} = 0 ,
	\label{eq:dRN1}
\end{equation}
where $\tau=n,p$.
We also have to constrain $\Delta _{\perp i}^{(n)} {\cal R}$ in a similar way
\begin{eqnarray}
	\label{eq:dQ3}
	\Delta _{\perp i}^{(n)} {\cal R} \cdot f^{(n)} &=& 0 ,\\
	\label{eq:dRN2}
	\Delta _{\perp i}^{(n)} {\cal R} \cdot N^{20(n)}_{i \tau} &=& 0. 
\end{eqnarray}
The six conditions (Eqs.~(\ref{eq:dQ1}) and~(\ref{eq:dRN1}--\ref{eq:dRN2})) 
give a set of equations which can be solved for the 
six parameters $\epsilon_i^{(n)}$, $\eta_{i \tau}^{(n)}$, $\lambda_i^{(n)}$ 
and $\mu_{i \tau}^{(n)}$ for each $i$ and $n$. The expressions for $\epsilon$,
$\eta$, $\lambda$ and $\mu$ can be found in the appendix. 

The quality of the collective coordinate found above can be quantified and 
calculated. This is 
done by calculating the decoupling measure, $D$, derived in~\cite{DK00}. One 
way of calculating $D$ is by computing the inverse mass matrix $ B_{\alpha \beta}$
 and then calculate
\begin{equation}
	\label{eq:D1}
	\breve{B}_{11} = \frac{d \xi^\alpha}{dQ} B_{\alpha \beta} 
		\frac{d \xi^\beta}{dQ} ,
\end{equation}
where we can approximate the derivative by the finite differences
\begin{equation}
	\label{eq:D2}
	\frac{d \xi^\alpha}{dQ} = \sqrt{2} \frac{\Delta {\cal R}^\alpha}{\Delta Q} .
\end{equation}
The decoupling measure is then calculated to be
\begin{equation}
	\label{eq:D3}
	D = \breve{B}_{11}-1,
\end{equation}
where we have uses the normalisation Eq.~(\ref{eq:CMass}). This
quantity is straightforward to evaluate, but it is easier to
understand from an alternative expression for $D$, which is based on
the generalisation of Eq.~(\ref{eq:dQ1}) to all coordinates
\begin{equation}
	\label{eq:D4}
	\Delta q^\mu = \sqrt{2} \Delta {\cal R} \cdot f^\mu.
\end{equation}
This leads to
\begin{equation}
	\label{eq:D5}
	D = \sum_{\mu>1}\left( \frac{\Delta q^\mu}{\Delta Q} \right)^2 ,
\end{equation}
i.e., $D$ is the sum of squares of the change of the non-collective coordinates with
the collective coordinate. This is clearly zero for exact decoupling.

\subsection{Projection basis for the LHA\label{sec:pLHA}}

One of the main difficulties of applying the LHA method to realistic
nuclear problems is the effort required in diagonalising the
large-dimensional RPA matrix repeatedly within the double iterative
process. To limit the computational effort we use the method
presented in Ref.~\cite{NW99} to reduce the size of the RPA matrix.
There it was shown that the RPA equation can be solved with good
accuracy by assuming that the RPA eigenvectors can be described as a
linear combination of a small number of state-dependent one-body
operators. The quality of the results, and the number of operators
needed depends strongly on the choice of the set of operators.
How to choose these operators is a longstanding problem in nuclear 
physics~\cite{Ho68,NK02}. 

We select a small number of one-body operators $F^{(k)}$, $k=1,\ldots,n$,
assuming that the RPA eigenvectors can be approximated as linear combinations 
of the $F^{(k)}$. The approximate RPA vector $\bar{f}_{,\alpha}$ is then given by 
\begin{equation}
	f_{,\alpha} \approx \bar{f}_{,\alpha} = \sum_{k=1}^{n} c_k 
		{\cal F}^{(k)}_{,\alpha}
	\label{eq:f1}
\end{equation}
where ${\cal F}^{(k)}$ is the expectation value of $F^{(k)}$.
To determine the coefficients $c_k$ the RPA matrices are projected onto 
the subspace $\{{\cal F}^{(k)}_{,\alpha}\}$:
\begin{eqnarray}
	\label{eq:M1}
	{\bf M}^{kl} &=& {\cal F}^{(k)}_{,\alpha} B^{\alpha \beta} 
	V_{;\beta \gamma} B^{\gamma \delta} {\cal F}^{(l)}_{,\delta} , \\
	\label{eq:N1}
	{\bf N}^{kl} &=& {\cal F}^{(k)}_{,\alpha} B^{\alpha \beta} 
	{\cal F}^{(l)}_{,\beta}.
\end{eqnarray}
The RPA equation can then be expressed as
\begin{equation}
	{\bf M}^{kl} c_l = \left( \hbar \bar{\Omega} \right)^2 {\bf N}^{kl} c_l ,
	\label{eq:pRPA1}
\end{equation}
where $\hbar \bar{\Omega}$ is a eigenfrequency of the projected RPA.
The rank of the matrix we need to diagonalise to solve the RPA problem has 
been reduced from the number of 2-quasi-particle 
degrees of freedom to the number of one-body operators  chosen.

\subsection{Schr\"{o}dinger equation on the collective path}
\label{sec:SE}

After having made a semi-classical approximation, which leads to a
classical Hamiltonian, we need to remember that we are studying a
quantum system. The standard technique to deal with this is to treat the
classical Hamiltonian as a quantum one, and to calculate the
eigenfunctions and energies. This is superficially similar to the generator
coordinate method, especially in the Gaussian overlap approximation
\cite{RS80}, but it is actually rather different. The key point is
the appearance of the kinetic terms, which correspond to time-odd
generator coordinates (usually not include in the GCM).

As discussed in Ref.~\cite{DK00}, we can include all manner of quantum
corrections to the potential energy, especially if we are interested
in \emph{absolute} values of the energy eigenvalues. On the other hand,
shape mixing---a spread of the wave function along the collective
path---is rather insensitive to these quantum corrections. Therefore, we
shall consider the Hamiltonian along the collective path without
further quantum corrections.

One must include the zero modes when quantising the Hamiltonian, since
they describe rotational and other excitations. Quantisation of the
Hamiltonian in a metric coordinate space turns the kinetic energy into a
Laplace-Beltrami operator (see, e.g., Ref.~\cite{AM88}) in the
relevant space,
\begin{equation}
	\label{eq:OB1}
	H(\vec{ X}) = - \frac{\Delta_g}{2}  + V(\vec{ X}) .
\end{equation}
The collective
Schr\"{o}dinger equation can then be written as
\begin{equation}
	\label{eq:SE1}
	H(\vec{ X}) \Psi(\vec{ X}) = E \Psi(\vec{ X}).
\end{equation}
In this paper we discuss calculations with one true collective
coordinate, and a number of additional momenta for the zero modes
(denoted as $p^{\text{ZM}}_i$): two or three angular momenta, depending
on whether the state is axial or not, and two operators connected to a
change of phase of the proton and neutron pairing gap, associated with
pairing rotation. These latter quantise as $\frac{1}{i}\frac{\partial
}{\partial \phi_N}$ and $\frac{1}{i}\frac{\partial }{\partial
\phi_P}$.

The potential $V$ is invariant under all the zero-modes, 
and only depends on the collective parameter,
\begin{equation}
	\label{eq:potential1}
	V(\vec{ X}) = V(Q) .
\end{equation}
The Laplace-Beltrami operator, $\Delta_g$, with variable (but
diagonal) mass matrix , where the zero-mode masses are given by
$B_i=B_i(Q)$, can then be written as
\begin{equation}
	\label{eq:LB1}
	\Delta_g = g^{-1/2} \frac{\partial}{\partial Q} \left( \frac{1}{B_Q}
	g^{1/2} \frac{\partial }{\partial Q} \right) -
	\sum_i g^{-1/2} p^\text{ZM}_i \left( \frac{1}{B_{i}} g^{1/2}p^\text{ZM}_i \right) 
\end{equation}
where $g=B_Q \prod_i B_i$, and $B_Q$ is identical to $B$ in
Eq.~(\ref{eq:CMass}), and thus equals 1. Below we shall write
$B_{\phi_N,\phi_P}$ for the neutron and proton pairing-rotational
masses. These are calculated as
\begin{eqnarray}
	\label{eq:BN}
	B_{\phi_N} &=& N^{20}_{n \alpha} V^{\alpha \beta} N^{20}_{n \beta}, \\
	\label{eq:BP}
	B_{\phi_P} &=& N^{20}_{p \alpha} V^{\alpha \beta} N^{20}_{p \beta},
\end{eqnarray}
and the rotational moments of inertia are defined in the usual way.

Since the potential and the masses are independent of the zero-mode coordinates
the wave-function $\Psi$ can be separated into various pieces,
\begin{equation}
	\label{eq:wf1}
	\Psi(Q,\phi_N,\phi_P,\Omega) =  g^{-1/4} U(Q) \frac{1}{\sqrt{2 \pi}}e^{i m \phi_N} 
	\frac{1}{\sqrt{2 \pi}}e^{i k \phi_P} D^I_{MK}(\Omega)^*
\end{equation}
where $m$ and $k$ are the quantum numbers for neutron and proton
pairing rotation, and $I,M,K$ are the usual rotator quantum
numbers. We shall be looking at ground states (bandheads) only, and
therefore we shall now use $I=M=K=0$, and since pairing rotational excitation corresponds to a change
in particle number, we shall be use $m=k=0$ as well. 
 Equation~(\ref{eq:LB1}) acting on $\Psi$ can now be rewritten as
\begin{eqnarray}
	g^{1/4} \Delta_g 2 \pi \Psi &=& 
	  g^{-1/4}\frac{\partial}{\partial Q}  \left(  g^{1/2}
	\frac{\partial }{\partial Q}g^{-1/4}U(Q) \right) \nonumber\\
	&=& U''(Q)+\frac{3{g'}^2-4g g''}{16g^2}U(Q) .
	\label{eq:LB2}
\end{eqnarray}
Here we see the typical reason to absorb the factor $g^{1/4}$ into the
wave function: it removes the linear derivative term, and we obtain a
``centrifugal'' potential in its place. This is fully consistent with
the standard procedure for separation of variables in radially
symmetric problems in two and three dimensions as can be found in any
quantum mechanics textbook.  Using (\ref{eq:wf1}) to separate
variables, the Schr\"{o}dinger equation~(\ref{eq:SE1}) can be written as
\begin{equation}
	\label{eq:SE2}
	-g^{-1/4} \frac{\partial}{\partial Q} 
	\left( g^{1/2} \frac{\partial }{\partial Q} g^{-1/4}U(Q)\right) 
	+V(Q) U(Q)
	= E U(Q) .
\end{equation}

Since we wish the wave function $\Psi$ to be normalisable we require
it to be finite, and we must then insist that $U(Q)$ goes to zero
when $g$ does. In the present work that only occurs when either of the
pairing gaps collapses and thus $B_{\phi_{P,N}}=0$ is zero, and we
shall ignore the rotational moments of inertia, which do not change
very quickly.  Below we solve Eq.~(\ref{eq:SE2}) on a grid with the
boundary condition that $U(Q_\text{max})=U(Q_\text{min})=0$. At points
where $B_\phi=0$ the condition $U=0$ holds exactly; for other cases
applying this boundary condition will only give an upper limit on
energy. 

The scaling of the wave-function removes $g$ from expectation values, and
the expectation value of any local operator $A(Q)$ can evaluated as
\begin{equation}
	\label{eq:evInt1}
	\left< A \right> = \int U(Q) A(Q) U(Q)  dQ 
\end{equation}
which shows that $U$ must be  normalised according to 
\begin{equation}
	\label{eq:evInt2}
	\int U(Q)^2  dQ = 1 . 
\end{equation}

\section{Results\label{sec:Results}}
To test the projection basis discussed in section~\ref{sec:pLHA} we implement
our method for a interaction and configuration space that where the 
approximation can be compared with exact results. 
\subsection{Pairing+quadrupole model}
We apply the LHA to the pairing+quadrupole Hamiltonian as described 
in~\cite{BK68}. With a constraint on both neutron and proton numbers the Hamiltonian can be 
written as
\begin{eqnarray}
	\label{eq:HPQ1}
	H' &=& H - \sum_{\tau=n,p} \mu_\tau N_\tau , \\
	\label{eq:HPQ2}
	H &=& \sum_k \epsilon_k c^\dagger_k c_k - \sum_{\tau=n,p} 
	\frac{G_\tau}{2} \left( P^\dagger_\tau P_\tau + P_\tau P^\dagger_\tau 
	\right) - \frac{\kappa}{2} \sum_{M=-2}^2 Q^\dagger_{2M} Q_{2M} , 
\end{eqnarray}
where $\epsilon_k$ are spherical single-particle energies, $N_\tau$ is the 
particle number operator, $Q_{2M}$ is the dimensionless quadrupole operator
\begin{equation}
	\label{eq:Q1}
	Q_{2M} = \frac{1}{\sqrt{2} b^2_0} \sum_{kl} \left< k \right| r^2 Y_{2M}
	\left| l \right> c^\dagger_k c_l ,
\end{equation}
where $b_0=1/\sqrt{\omega_0}$ is the standard oscillator length
and $P^\dagger_\tau$ is the (dimensionless) pairing operator
\begin{equation}
	\label{eq:P1}
	P^\dagger_\tau = \sum_{k>0} c^\dagger_k c^\dagger_{-k}.
\end{equation}
This Hamiltonian is treated in the Hartree-Bogoliubov approximation and it 
has been shown that at the minimum the local RPA for this Hamiltonian is 
equivalent to the quasi-particle RPA.
For $M=1, 2$ we rewrite the quadrupole operators of Eq.~(\ref{eq:Q1}) 
as sums and differences, 
\begin{equation}
	\label{eq:Q2}
	Q^{(\pm)}_{2M} = \frac{1}{\sqrt{2}} \left( Q_{2M} \pm Q_{2-M} \right),\quad (M=1,2)\quad,
\end{equation}
and the pairing operator of Eq.~(\ref{eq:P1}) as 
\begin{equation}
	\label{eq:P2}
	P^{(\pm)}_\tau = \frac{1}{\sqrt{2}} \left(P_\tau \pm 
	P^\dagger_\tau \right), \qquad \tau=p,n.
\end{equation}
The pairing and quadrupole operators can then be arranged into five Hermitian,
$R_i$, and four anti-Hermitian, $S_j$, operators:
\begin{eqnarray}
	\label{eq:PQ1}
	R_i &=& \left( P^{(+)}_n, P^{(+)}_p, Q_{20}, Q^{(-)}_{21}, Q^{(+)}_{22} 
	\right) , \\ 
	\label{eq:PQ2}
	S_j &=& \left( P^{(-)}_n, P^{(-)}_p, Q^{(+)}_{21}, Q^{(-)}_{22} \right).
\end{eqnarray}
The Hamiltonian of Eq.~(\ref{eq:HPQ2}) can then be written as
\begin{equation}
	\label{eq:HPQ3}
	H = \sum_k \epsilon_k c^\dagger_k c_k - \frac{1}{2} \sum_i \kappa_i 
	R_i R_i + \frac{1}{2} \sum_j \kappa_j S_j S_j , 
\end{equation}
with $\kappa_{i(j)}= G_\tau$ for $P^{(\pm)}_\tau$ and $\kappa_{i(j)}= \kappa$ for the 
$Q$ operators.
After solving the mean-field problem within the Hartree-Bogoliubov approximation
the mass matrix and RPA potential around the minimum can be calculated as
\begin{eqnarray}
	\label{eq:B1}
	B^{\alpha \beta} &=& E_\alpha \delta_{\alpha \beta} - 2 \sum_j
	\kappa_j (S_j)_\alpha (S_j)_\beta , \\
	\label{eq:V1}
	V_{;\alpha \beta} &=& E_\alpha \delta_{\alpha \beta} - 2 
	\sum_i \kappa_i (R_i)_\alpha (R_i)_\beta,
\end{eqnarray}
where $E_\alpha=e_q+e_{q'}$ is the 2 quasi-particle energy and $\alpha$ and $\beta$ 
label 2 quasi-particle states. $O_\alpha$ is the 2 quasi-particle matrix element of the 
operator $O$, which can also be written as $O^{20}_{qq'}$ with $\alpha=qq'$.

The spherical single particle energies are taken from~\cite{BK68}. Our 
model space consists of two major shells. We follow~\cite{BK68} and
multiply all quadrupole matrix elements with the factor 
\begin{equation}
	\label{eq:Qfactor1}
	\zeta = \frac{N_L+\frac{3}{2}}{N_H+\frac{3}{2}},
\end{equation}	
where $N_L$ is the harmonic oscillator quantum number of the lower major shell and $N_H$ 
that of the higher one. To achieve the 
same root-mean-square radii for protons and neutrons different harmonic 
oscillator frequencies are adopted for each type of nucleons. As a result the 
proton and neutron quadrupole operators are multiplied by the factors
\begin{equation}
	\label{eq:Qfactor2}
	\alpha_n = \sqrt{ \frac{2N}{A} }  \qquad {\rm and} \qquad
	\alpha_p = \sqrt{ \frac{2Z}{A} }
\end{equation} 
where $N$ ($Z$) is the neutron (proton) number and $A=N+Z$.

We have chosen a set of representative isotopes for the examples shown in this 
paper. For $^{54}$Cr, $^{58}$Fe and $^{62}$Zn the interaction strengths are 
chosen to reproduce the ground state deformation listed in~\cite{MN95} and the 
pairing strengths are chosen to approximately reproduce the relation give 
in~\cite{MN92}. The interaction strengths for the isotopes $^{66}$Zn and 
$^{70}$Zn are 
chosen to be the same as for $^{62}$Zn. The quadrupole- and pairing-strengths, 
$\kappa$ and $G_\tau$, are listed in table~\ref{tab:kapG} together with the 
corresponding deformations and pairing gaps.
\begin{table}%[htb] 
\caption{The quadrupole- and pairing interaction strengths, $\kappa$
and $G_\tau$, used for the examples discussed in this section. The
deformation and pairing gap calculated for those interaction
strengths are also listed.}
\label{tab:kapG}
\begin{ruledtabular}
\begin{tabular}{cddddddd}
 & \multicolumn{1}{c}{$\kappa$} [MeV]& \multicolumn{1}{c}{$G_n$} [MeV]& 
\multicolumn{1}{c}{$G_p$} [MeV]& \multicolumn{1}{c}{$\epsilon$} & 
\multicolumn{1}{c}{$\gamma$} & \multicolumn{1}{c}{$\Delta_n$} [MeV]& 
\multicolumn{1}{c}{$\Delta_p$} [MeV]\\
\hline
 $^{54}$Cr & 0.201367 & 0.525586 & 0.485390 & 0.167 & 0.0 & 1.60 & 1.60 \\
 $^{58}$Fe & 0.122188 & 0.379609 & 0.478824 & 0.213 & 0.0 & 1.60 & 1.60 \\
 $^{62}$Zn & 0.113687 & 0.375498 & 0.411877 & 0.192 & 0.0 & 1.63 & 1.70 \\
 $^{66}$Zn & 0.113687 & 0.375498 & 0.411877 & 0.286 & 60.0 & 1.95 & 1.74 \\
 $^{70}$Zn & 0.113687 & 0.375498 & 0.411877 & 0.772 & 0.0 & 0.63 & 0.96 \\
\end{tabular}
\end{ruledtabular}
\end{table}

\subsection{Improved approximate representation of the normal-mode operators}
\label{sec:improved}
The quality of the results achieved by the projection method described in 
section~\ref{sec:pLHA} strongly depends on the choice of the single particle 
operator basis. In Ref.~\cite{NW99} it was demonstrated  that a basis set
consisting of pairing, multipole and spin dependent one-body operators are 
not able to reproduce the results of a full RPA calculation. On the other hand 
if the basis is chosen to be a set of state-dependent Hermitian one-body 
operators of the structure
\begin{equation}
	\label{eq:basis1}
	\tilde{F}_k \equiv \sum_\alpha \frac{(F_k)_\alpha}{E_\alpha^2} 
	\left( a^\dagger a^\dagger \right)_\alpha + {\rm h.c.} 
\end{equation}
good agreement can be achieved with a small set of operators.
The suppression factor $E_\alpha^{-2} $ can be understood if one 
looks at a simple example~\cite{NW99}.
With the basis of
Eq.~(\ref{eq:basis1}) good results can be achieved for the low-lying 
$\beta$- and $\gamma$-vibrations~\cite{NW99}, as can be seen in 
table~\ref{tab:bgpvib1}, with a small set of operators 
consisting of the 8 pairing and quadrupole operators
\begin{equation}
	\label{eq:basis2}
	\tilde{F}_k = \left( \tilde{P}^{(+)}_\tau, \tilde{P}^{(-)}_\tau, 
	\tilde{Q}_{20 \tau}, \tilde{Q}^{(+)}_{22 \tau} \right) , \qquad \tau=n,p .
\end{equation}
Even though the $\beta$- and $\gamma$-vibrations are well described with this
basis set, the higher lying solutions of pairing-vibrational character are not 
well described.  A couple of examples are listed in table~\ref{tab:bgpvib1}. 
These results are 
not significantly improved by including higher-order multipole or 
quadrupole-pairing operators in the basis. 
\begin{table}%[htb] 
\caption{Comparing the full RPA energy, $\hbar \Omega$, the projected
RPA energy, $\hbar \bar{\Omega}$, and $\delta_{B,1}$ for the
$\beta$-, $\gamma$-, $\Delta^{(1)}$- and $\Delta^{(2)}$-vibrations
using the projection basis of~\cite{NW99}.  The energies are in units of
MeV.}
\label{tab:bgpvib1}
\begin{ruledtabular}
\renewcommand{\arraystretch}{1.2}
\begin{tabular}{cdddd}
 & \multicolumn{4}{c}{$\beta$-vibration} \\
\cline{2-5}
 & \multicolumn{1}{c}{$\hbar \Omega$} & \multicolumn{1}{c}{$\hbar \bar{\Omega}$} &
 \multicolumn{1}{c}{$\delta_B$} & \multicolumn{1}{c}{$\delta_1$} \\
\cline{2-5} 
 $^{54}$Cr & 1.216 & 1.246 & 0.0013 & 0.6528 \\
 $^{58}$Fe & 2.511 & 2.656 & 0.0282 & 0.0910 \\
 $^{62}$Zn & 1.966 & 2.033 & 0.0053 & 0.3593 \\
 $^{66}$Zn & 2.117 & 2.188 & 0.0050 & 0.3117 \\
 $^{70}$Zn & 1.029 & 1.068 & 0.0014 & 0.2146 \\
\cline{2-5}
& \multicolumn{4}{c}{$\gamma$-vibration} \\
\cline{2-5}
& 
\multicolumn{1}{c}{$\hbar \Omega$} & \multicolumn{1}{c}{$\hbar \bar{\Omega}$} &
 \multicolumn{1}{c}{$\delta_B$} & \multicolumn{1}{c}{$\delta_1$} \\
 $^{54}$Cr & 2.289 & 2.386 & 0.0103 & 0.0085 \\
 $^{58}$Fe & 1.971 & 2.049 & 0.0090 & 0.0076 \\
 $^{62}$Zn & 1.284 & 1.298 & 0.0010 & 0.0010 \\
 $^{66}$Zn & 1.280 & 1.289 & 0.0003 & 0.0002 \\
 $^{70}$Zn & 3.166 & 3.348 & 0.0563 & 0.0511 \\
\cline{2-5}
 & \multicolumn{4}{c}{$\Delta^{(1)}$-vibration} \\
\cline{2-5}
 & \multicolumn{1}{c}{$\hbar \Omega$} & \multicolumn{1}{c}{$\hbar \bar{\Omega}$} &
 \multicolumn{1}{c}{$\delta_B$} & \multicolumn{1}{c}{$\delta_1$} \\
\cline{2-5}
 $^{54}$Cr & 3.208 & 3.894 & 0.1591 & 0.1496\\
 $^{58}$Fe & 3.239 & 3.935 & 0.2712 & 0.2534\\
 $^{62}$Zn & 3.383 & 4.173 & 0.5220 & 0.5107\\
 $^{66}$Zn & 3.661 & 5.161 & 0.5171 & 0.4618\\
 $^{70}$Zn & 1.814 & 2.052 & 0.0085 & 0.0105\\
\cline{2-5}
& 
\multicolumn{4}{c}{$\Delta^{(2)}$-vibration} \\
\cline{2-5}
& 
\multicolumn{1}{c}{$\hbar \Omega$} & \multicolumn{1}{c}{$\hbar \bar{\Omega}$} &
 \multicolumn{1}{c}{$\delta_B$} & \multicolumn{1}{c}{$\delta_1$} \\
\cline{2-5}
 $^{54}$Cr & 3.449 & 4.505 & 0.5237 & 0.5113 \\
 $^{58}$Fe & 3.549 & 4.824 & 0.6442 & 0.6060 \\
 $^{62}$Zn & 3.537 & 4.797 & 0.7934 & 0.7751 \\
 $^{66}$Zn & 4.255 & 5.493 & 0.7827 & 0.7977 \\
 $^{70}$Zn & 3.573 & 4.824 & 0.6801 & 0.6809 \\
\end{tabular}
\end{ruledtabular}
\end{table}

To improve the results for the pairing vibrations we include a pairing operator 
only active close to the Fermi-surface. To avoid the problem of 
having to select by hand which levels that would have a non-zero matrix element we simply divide the 
standard pairing operator $P_\pm$ with a large power of $E_\alpha$. If the 
suppression factor, $E_\alpha^k$, is chosen with a large enough $k$ all matrix 
elements except the ones with $E_\alpha$ close to zero will become negligible 
and the result will not depend on $k$. The basis set is now
\begin{equation}
	\label{eq:basis3}
	\tilde{F}_k = \left( \tilde{P}^{(+)}_\tau, \tilde{P}^{(-)}_\tau, 
	\tilde{Q}_{20 \tau}, \tilde{Q}^{(+)}_{22 \tau}, 
	\frac{\tilde{P}^{(+)}_\tau}{E_\alpha^k}, 
	\frac{\tilde{P}^{(-)}_\tau}{E_\alpha^k} 
	\right) , \qquad \tau=n,p .
\end{equation} 
We have chosen $k=10$. From table~\ref{tab:bgpvib2} we can
 see that almost all the low-lying vibrational modes are now described with a
 very high accuracy. 

\begin{table}%[htb]
\caption{Comparing the full RPA energy, $\hbar \Omega$, the projected
RPA energy, $\hbar \bar{\Omega}$, and  $\delta_{B,1}$ for the
$\beta$-, $\gamma$-, $\Delta^{(1)}$- and $\Delta^{(2)}$-vibrations
using the new projection basis (\ref{eq:basis3}).  The energies are in units of MeV. }
\label{tab:bgpvib2}
\begin{ruledtabular}
\renewcommand{\arraystretch}{1.2}
\begin{tabular}{cdddd}
 & \multicolumn{4}{c}{$\beta$-vibration} \\
\cline{2-5}
 & \multicolumn{1}{c}{$\hbar \Omega$} & \multicolumn{1}{c}{$\hbar \bar{\Omega}$} &
 \multicolumn{1}{c}{$\delta_B$} & \multicolumn{1}{c}{$\delta_1$} \\
\cline{2-5}
 $^{54}$Cr & 1.216 & 1.225 & 0.0003 & 0.6633\\
 $^{58}$Fe & 2.511 & 2.537 & 0.0030 & 0.0630\\
 $^{62}$Zn & 1.966 & 2.004 & 0.0023 & 0.3613\\
 $^{66}$Zn & 2.116 & 2.131 & 0.0010 & 0.3156\\
 $^{70}$Zn & 1.029 & 1.031 & 0.0001 & 0.2110\\
\cline{2-5}
& \multicolumn{4}{c}{$\gamma$-vibration} \\
\cline{2-5}
& 
\multicolumn{1}{c}{$\hbar \Omega$} & \multicolumn{1}{c}{$\hbar \bar{\Omega}$} &
 \multicolumn{1}{c}{$\delta_B$} & \multicolumn{1}{c}{$\delta_1$} \\
\cline{2-5}
 $^{54}$Cr & 2.289 & 2.386 & 0.0103 & 0.0085 \\
 $^{58}$Fe & 1.971 & 2.049 & 0.0090 & 0.0076 \\
 $^{62}$Zn & 1.284 & 1.298 & 0.0010 & 0.0010 \\
 $^{66}$Zn & 1.280 & 1.289 & 0.0003 & 0.0002 \\
 $^{70}$Zn & 3.166 & 3.348 & 0.0363 & 0.0311 \\
\cline{2-5}
 & \multicolumn{4}{c}{$\Delta^{(1)}$-vibration} \\
\cline{2-5}
 & \multicolumn{1}{c}{$\hbar \Omega$} & \multicolumn{1}{c}{$\hbar \bar{\Omega}$} &
 \multicolumn{1}{c}{$\delta_B$} & \multicolumn{1}{c}{$\delta_1$} \\
\cline{2-5}
 $^{54}$Cr & 3.208 & 3.212 & 0.0008 & 0.0022\\
 $^{58}$Fe & 3.239 & 3.247 & 0.0021 & 0.0039\\
 $^{62}$Zn & 3.383 & 3.392 & 0.0055 & 0.0099\\
 $^{66}$Zn & 3.661 & 3.664 & 0.0005 & 0.0037\\
 $^{70}$Zn & 1.814 & 1.815 & 0.0001 & 0.0088\\
\cline{2-5}
& 
\multicolumn{4}{c}{$\Delta^{(2)}$-vibration} \\
\cline{2-5}
 & 
\multicolumn{1}{c}{$\hbar \Omega$} & \multicolumn{1}{c}{$\hbar \bar{\Omega}$} & 
\multicolumn{1}{c}{$\delta_B$} & \multicolumn{1}{c}{$\delta_1$}\\
\cline{2-5}
 $^{54}$Cr & 3.449 & 3.596 & 0.0426 & 0.0370 \\
 $^{58}$Fe & 3.549 & 3.647 & 0.0176 & 0.0309 \\
 $^{62}$Zn & 3.537 & 3.621 & 0.0210 & 0.0201 \\
 $^{66}$Zn & 4.254 & 4.767 & 0.9355 & 0.9286 \\
 $^{70}$Zn & 3.573 & 4.039 & 0.1649 & 0.1855 \\
\end{tabular}
\end{ruledtabular}
\end{table}
To check if the wave functions are described as well as the RPA energies 
by the projection method we also calculate the overlap of the full RPA vector $f$
and the projected RPA vector $\bar{f}$. As criteria for good projection we use 
smallness of the following quantities
\begin{eqnarray}
	\label{eq:deltaB}
	\delta_B = 1-f_{,\alpha} B^{\alpha \beta} \bar{f}_{,\beta} , \\
	\label{eq:delta1a}
	\delta_1 = 1 - \frac{\left( f,\bar{f} \right)}{\sqrt{\left(f,f \right) 
	\left( \bar{f}, \bar{f} \right)}} ,
\end{eqnarray}
where 
\begin{equation}
	\label{eq:delta1b}
	\left(f,f' \right) = \sum_\alpha f_{,\alpha} f'_{,\alpha} .
\end{equation}
If $\delta=0$ the projection corresponds to an exact result. The
difference between $\delta_B$ and $\delta_1$ is that an admixture of a
spurious solution will contribute to $\delta_1$ and not $\delta_B$;
$\delta_B$ is the consistent quantity from the topological analysis.
In table~\ref{tab:bgpvib1} we can see that $\delta$ has a relative
small value for the $\beta$- and $\gamma$-vibration but, as expected,
a substantially larger value for the pairing-vibrations. The new
projection basis does systematically improve the wave-function as well
as the energy, as can be seen in table~\ref{tab:bgpvib2} where the
values of $\delta$ are much smaller. The exception being the second pairing vibration in $^{66}$Zn. 
This is due to that the ordering of the RPA solutions in the full RPA relative to the 
projected RPA is different in this case. Since the improved basis for the
projected RPA gives energy spectra and wave-functions that are much
better than the set used in Ref.~\cite{NW99}, we will use the new set
in the following calculations.

Calculating the collective path using the projection basis has an advantage
besides reducing the rank of the RPA matrix. We do not have 
any spurious solutions in the projected RPA calculations since we have 
not included angular momentum and particle number operators in the basis.
We can therefore avoid problems due to crossings between the spurious modes 
and the physical modes along the collective path. Away from the minimum there 
is still an admixture of the spurious solution into the collective coordinate. 
[As stated above, we should use a symplectic RPA to resolve this problem, 
but will not do so here due to its complexity.]

\subsection{Representative case for large amplitude collective motion} 
We would like to perform as simple a test of our method as possible,
especially, we would like the model space to be small.  We decided to
concentrate on $^{58}$Fe; since this nucleus is $\gamma$-soft, it
should provide a demanding testing ground for our methodology.  We
have calculated the collective path using both the full RPA and the
projected RPA.

\begin{figure}%[htbp]
\centerline{\includegraphics[clip,width=12cm]{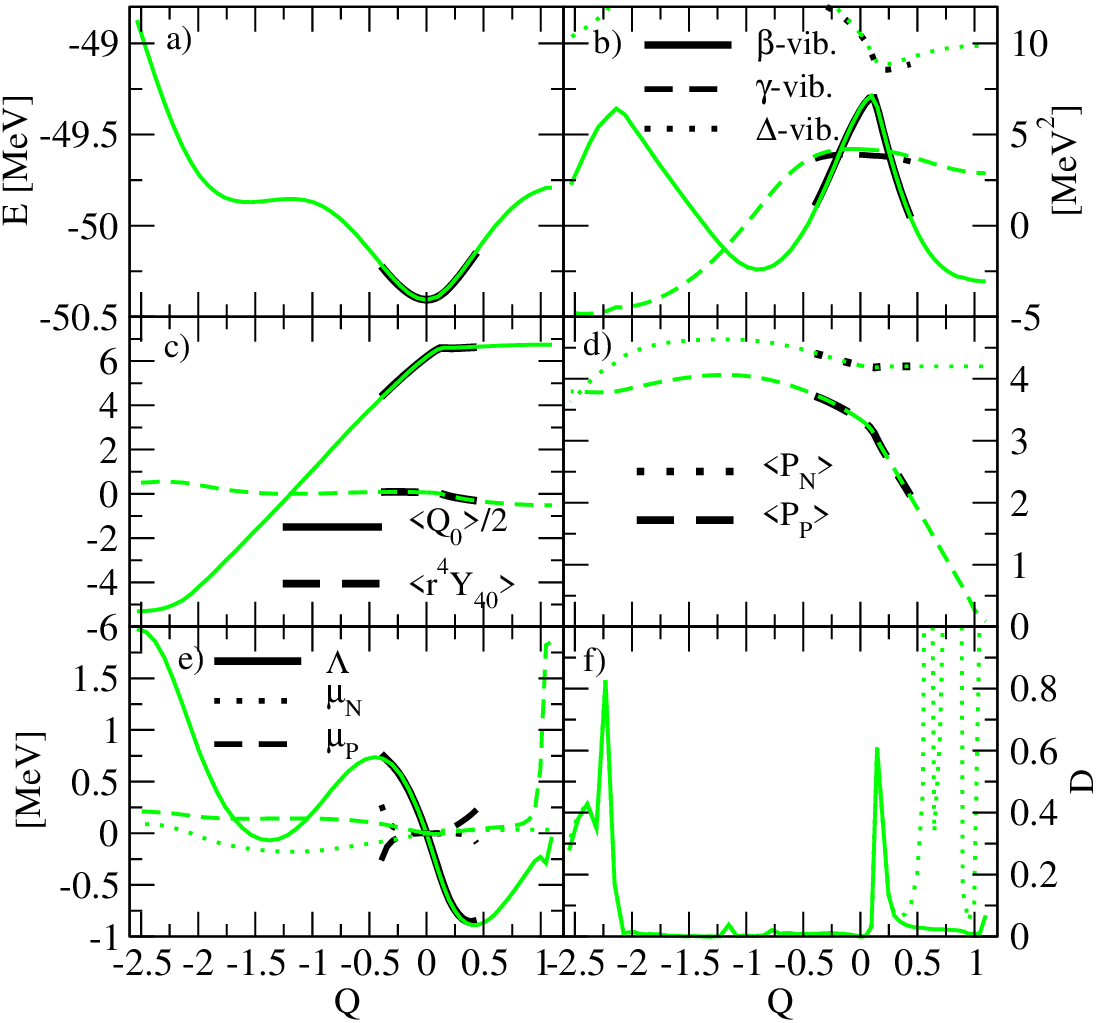}}
%  \centerline{\psfig{clip=true,figure=fe58beta.eps,height=13cm,angle=0}} 
  \caption{(Color online) Large amplitude collective motion in $^{58}$Fe with axial
  symmetry (following the $\beta$-vibration, the second lowest RPA
  solution at equilibrium). a) Energy along the collective path.  b)
  The square of the lowest RPA frequencies.  c) The dimensionless
  quadrupole moment, $\left< Q_0 \right>$, and the hexadecapole
  moment, $\left< r^4 Y_{40} \right>$. d) The dimensionless pairing
  operators $\left< P_\tau \right>$.  e) The cranking parameter
  $\Lambda$ and the chemical potential $\mu$.  f) The decoupling
  measure, $D$. The dotted line represents the
  numerically over-complete decoupling measure and the solid line the
  correct calculation where the contribution due to over-completeness
  of the projection basis has been removed.  The grey (green online)
  curves represents the results for the projected RPA and the black
  curves are for the full RPA.}
  \label{fig:fe58beta}
\end{figure}

\begin{figure}%[bhtp]
  \centerline{\includegraphics[clip,width=6.0cm]{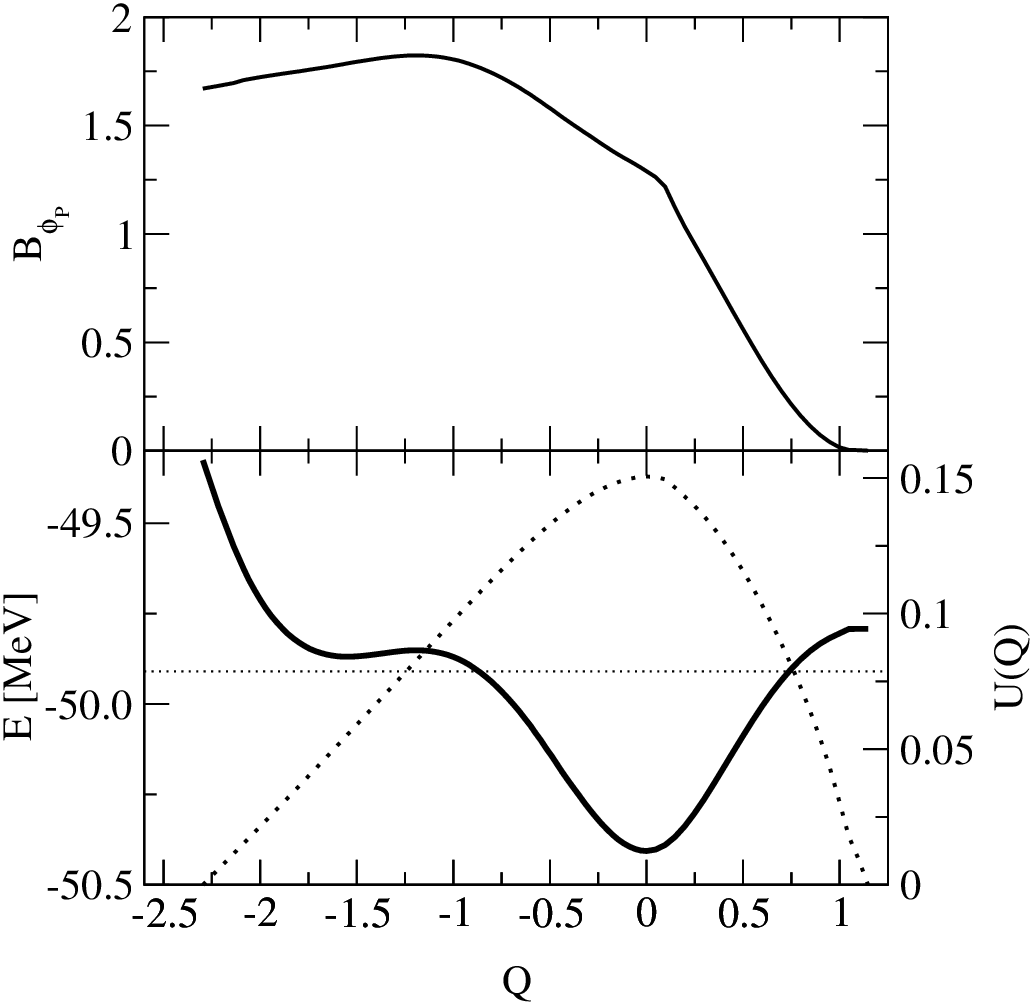}} 
  \caption{The upper panel show the proton rotational mass along the collective path for projected 
	axial large amplitude collective motion in $^{58}$Fe (following the 
	second RPA solution). The lower panel show energy and the radial wave 
	function for the large amplitude collective motion 
	in $^{58}$Fe, following the second RPA solution. The thick solid line 
	is the potential energy, the thin horizontal line give the position of
	the lowest eigenvalue. The energy scale is on the left side. The 
	wave function is shown as well (dotted line), with 
	the scale on the right side.} 
  \label{fig:SEfe58beta}
\end{figure}

\begin{figure}%[bhtp]
  \centerline{\includegraphics[clip,width=12cm,angle=0]{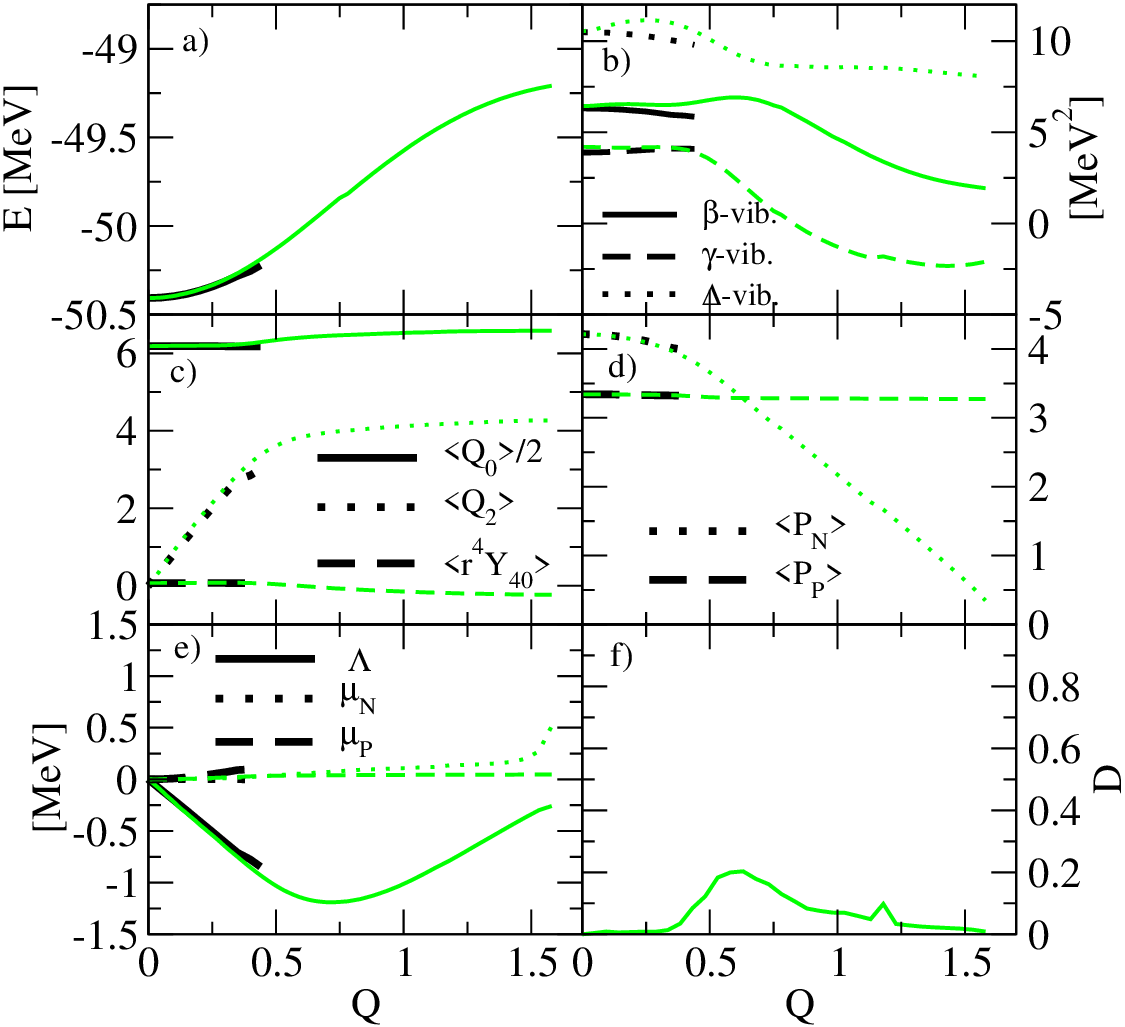}} 
%  \centerline{\psfig{clip=true,figure=fe58gam.eps,height=13cm,angle=0}} 
  \caption{(Color online) Large amplitude collective motion in $^{58}$Fe 
  (following the $\gamma$-vibration, the lowest RPA
  solution at equilibrium). a) Energy along the collective path. 
  b) The square of the lowest RPA frequencies.
  c) The dimensionless quadrupole moments, $\left<
  Q_0 \right>$, $\left<Q_2 \right>$ and the hexadecapole moment, $\left< r^4 Y_{40}
  \right>$. d) The dimensionless pairing operators $\left< P_\tau \right>$.
  e) The cranking parameter $\Lambda$ and the chemical potential $\mu$. 
  f) The decoupling measure, $D$. 
  The grey (green online) curves are the results for the projected
  RPA and the black curves are for the full RPA.} 
  \label{fig:fe58gam}
\end{figure}

\begin{figure}%[bhtp]
  \centerline{\includegraphics[clip,width=6.0cm]{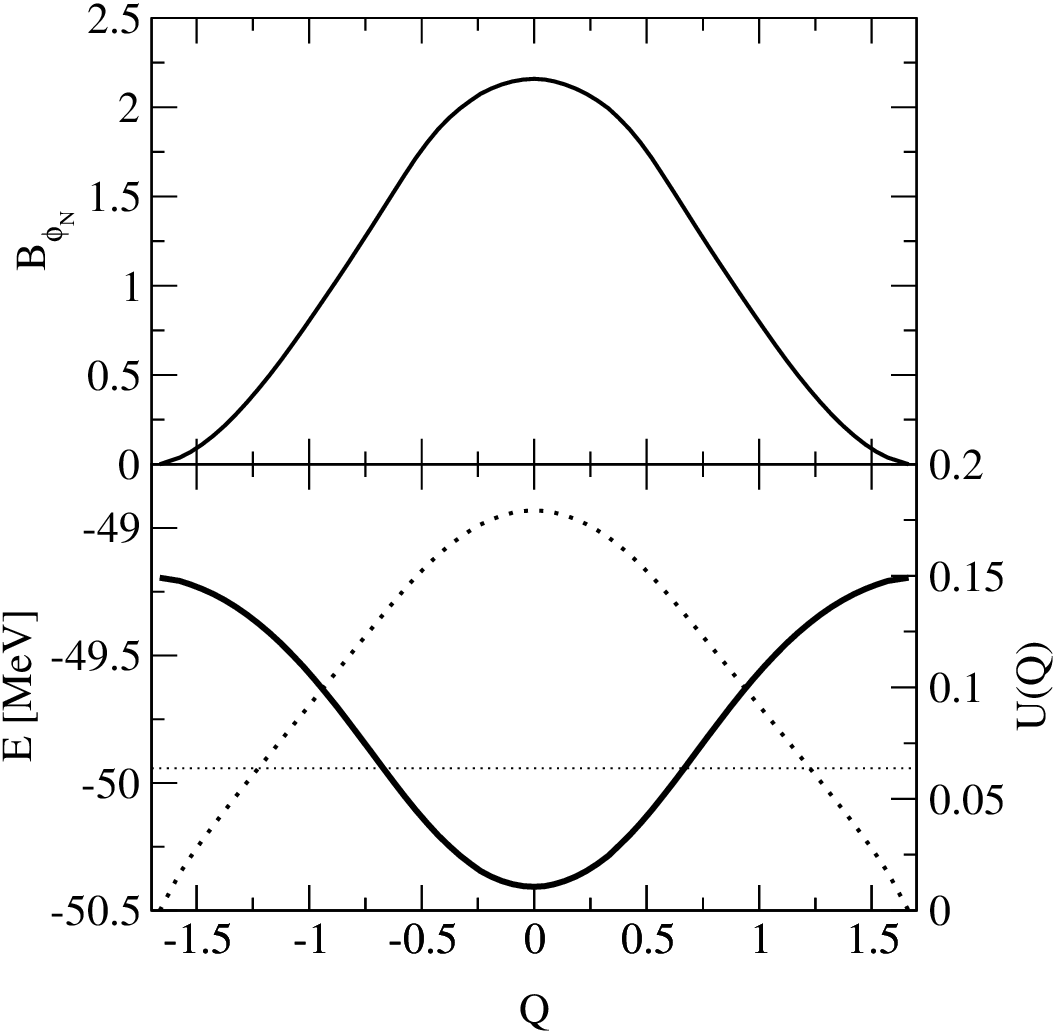}} 
  \caption{The upper panel show the proton rotational mass along the collective path for projected 
	large amplitude collective motion in $^{58}$Fe (following the 
	lowest RPA solution). The lower panel show energy and the radial wave 
	function for the large amplitude collective motion 
	in $^{58}$Fe, following the lowest RPA solution. The thick solid line 
	is the potential energy, the thin horizontal lines give the position of
	the lowest eigenvalue. The energy scale is on the left side. The 
	wave function is shown as well (dotted line), with 
	the scale on the right side.} 
  \label{fig:SEfe58gam}
\end{figure}

The results are shown in
Fig.~\ref{fig:fe58beta}--\ref{fig:SEfe58gam}. There is good agreement
between the projected and full RPA results along the collective path
in all cases.  This provides a further confirmation of the quality of
our projection basis and shows that the basis works well, also away
from the mean-field minimum. The collective path is found in a smaller
range of the collective coordinate when we are using the full RPA
compared to the results for the projected RPA. This is due to the fact
that the collective coordinate mixes with our spurious modes, which
leads to problem with convergence in our double iterative method, due
to the approximations made in the derivation. The mixing of the
collective coordinate and the spurious solution remains small as long
as the spurious mode is almost orthogonal to the collective
solution. When the energy of the spurious solution is similar to the
collective solution the denominator in the expression for the overlap
becomes small which causes the overlap to become large. In the
projected RPA calculation we do not have any spurious solutions since
we have not included any of the operators connected with the spurious
motion in our basis. Therefore we do not get a large spurious
contribution to our collective coordinate and we have better
numerical stability of our calculation.

We first investigate axial collective motion (see also~\cite{AW03}), by following the
$\beta$-vibration (even though this is not the lowest eigenvalue at
equilibrium, it is the lowest one of axial symmetry). From
Fig.~\ref{fig:fe58beta} we can see that the quadrupole moment is
approximately proportional to the collective coordinate $Q$ in the
region $-2 < Q <0$, which is an indication that we have a
path relative close to what we would obtain from a mean-field
calculation with a constraint on the quadrupole moment. At larger and
smaller values of $Q$ the deformation $\left<Q_0 \right>$ remains
almost constant. Instead, the collective coordinate is now dependent
on the pairing fields, for large $Q$ proton pairing and for small $Q$
neutron pairing. At $Q \approx 1.1$ the proton pair field collapses to
zero. Our collective path ends at this point, since the singularity at
zero pairing is similar to the origin in polar coordinates, with $Q$
playing the role of radial coordinate and the pairing phase $\phi$ the
role of polar angle. The change
from quadrupole to pairing mode is dominated by a narrowly avoided crossing
with the lowest pairing-vibration at $Q \approx 0.2$.  After this
crossing the quadrupole moment, $\left< Q_0 \right>$ saturates and the
$ \left< P_p \right>$ starts changing. This avoided crossing shows that 
more then one collective coordinate would be needed for an accurate description of 
the collective dynamics. The RPA frequency of the
$\beta$-vibration is as expected proportional to the derivative of the
cranking parameter, $\Lambda$.

We have also looked at the potential energy, simply calculated as the expectation
value of the Hamiltonian at each point. In Fig.~\ref{fig:fe58beta} a)
we see that the potential has a local energy maximum at $Q\approx -1$,
which corresponds to a spherical shape, and a shallow oblate minimum
at $Q\approx -1.6$. The potential around the minimum show an a
quadratic behaviour which indicates that the harmonic approximation in
RPA is well fulfilled for small-amplitude collective motion, but
obviously fails for wave functions that have substantial support away
from the minimum. It can easily be seen that the regions where the
potential energy has a positive derivative are the regions where the
cranking parameter, $\Lambda$, has negative value and the converse.

The key to the whole approach is the decoupling parameter, $D$, which
is plotted in Fig.~\ref{fig:fe58beta} f). It has a small value
indicating a good decoupling of the collective mode from all the
non-collective modes. The exceptions are at $Q \approx 0.2$ and $Q <
-2$ which is due to two avoided crossings of the $\beta$-vibration
with the pairing vibrations, as can be seen in Fig.~\ref{fig:fe58beta}
b). The large peak in Fig.~\ref{fig:fe58beta} f) at $Q > 0.5$ is due
to an approximate numerical over-completeness in in the basis on which
we have projected the RPA vectors. The over-completeness comes when a
pair field is zero and the projected mass matrix has a zero
eigenvalue. This is due to the extra pairing term included in the
basis in section~\ref{sec:improved}.  The over-completeness appear
near the collapse of the proton pairing field, as can be seen in
Fig.~\ref{fig:fe58beta} d). Even tough the basis only becomes exactly
over-complete at the point of pairing collapse the calculation of $D$
is already influenced when the pair field is small, due to the fact
the $D$ is calculated from an inverse of the mass matrix (see
Eq.~(\ref{eq:D3})) which becomes ill defined.  We should of course
remove such a spurious contribution; this can be done quite easily,
and leads to the result plotted as a solid curve in
Fig.~\ref{fig:fe58beta} f). From now on we will only plot the value of
$D$ where the contribution from over-completeness has been
removed. The collapse of the pair field has a surprisingly strong
influence on the collective path. Whether this is a result of the 
approximations we made, our choice of force or a general feature is 
not clear at this point.

In section~\ref{sec:SE} it was described how to solve the one-body Schr\"{o}dinger 
equation for the collective path. In the case discussed above the proton 
pair field collapses at $Q \approx 1.1$. We can therefore expect that proton 
pairing rotation will play a key role for the excitation spectrum of our system.
The proton pairing mass is plotted in Fig.~\ref{fig:SEfe58beta}. We can see 
that close to the collapse of the proton pairing 
$B_{\phi_P} \propto (Q-Q_{\rm max})^2$ which is what we expect when the collective 
coordinate is approximately $\left<P_P\right>$. For negative $Q$ $B_{\phi_P}$ has a 
non-trivial behaviour.
In Fig.~\ref{fig:SEfe58beta} we also show the lowest eigenvalue and radial wave function of the 
collective Hamiltonian including $k=0$ proton pairing rotation. At $Q=Q_{\rm max}$
where $B_{\phi_P}=0$ the ground-state wave-function goes to zero linear in 
$Q$ as expected from at the origin in polar coordinates. At $Q=Q_{\rm min}$ 
we have made the approximation that $U(Q_{\rm min})=0$. There is no bound state 
supported  by  the shallow oblate minimum and the lowest excited state is $1.30$ MeV 
above the collective ground state. This excitation energy is substantially smaller then the 
RPA harmonic approximation energy of $2.54$ MeV and reflects the anharmonic nature of the 
large-amplitude excitation.
In table~\ref{tab:fe58beta} we can see that the large component of the 
wave-function at small and negative $Q$ gives rise to a reduction of the 
expectation value of $Q_0$ by almost 30\% relative to the mean-field results.
\begin{table}%[htb] 
\caption{The expectation values of the quadrupole- and pairing operators for
	$k=0$ collective Hamiltonian along the axial collective path in
	$^{58}$Fe and at the mean-field minimum.}
\label{tab:fe58beta}
\begin{ruledtabular}
\begin{tabular}{cdd}
 & Collective & Mean-field  \\
\hline
 $\left<Q_0\right>$ & 8.98 & 12.37  \\
 $\left<P_N\right>$ & 4.36 & 4.21  \\
 $\left<P_P\right>$ & 3.25 & 3.34  \\
\end{tabular}
\end{ruledtabular}
\end{table}

In the case where we follow the path emerging from  the lowest mode, 
the $\gamma$-vibration, we can obtain similar results.
A number of results identifying the collective path are shown
in Fig.~\ref{fig:fe58gam}, where we have linear change of 
$\left< Q_2 \right>$ with the collective coordinate, while
all other expectation values remain relatively unchanged for $|Q| < 0.5$. At
larger values of the collective coordinate we see a saturation in 
$\left< Q_2 \right>$ and a strong reduction in the neutron pair field,
which finally collapses to zero. Once again, this is mediated by an avoided
crossing between quadrupole- and pairing-vibration modes.
The decoupling measure, $D$, in Fig.~\ref{fig:fe58gam} f) has a similar
behaviour as for the $\beta$-vibration. The crossing with the pairing
vibration is visible as an increase in $D$ at around $Q \approx 0.6$. 
At larger $Q$ we have a large
contributions to $D$ due to over-completeness of the basis this time
caused by the strongly reduced neutron pairing field. By mirroring the 
potential to negative $Q$ (and negative $\left<Q_2\right>$) we get a closed 
collective path from the neutron pairing collapse at $Q\approx 1.6$ to 
the mirrored neutron pairing collapse at $Q\approx -1.6$.
The neutron rotational pairing mass in Fig.~\ref{fig:SEfe58gam} shows the 
expected quadratic behaviour in $\left| Q \right|$ close to $Q_{\rm max}$. 
In Fig.~\ref{fig:SEfe58gam} we also show the eigenvalues and wave-functions of the 
collective Hamiltonian including $m=0$ neutron pairing rotation. The lowest excited state 
is at $1.49$ MeV above the collective ground state. This excitation energy is substantially 
smaller then the corresponding RPA excitation energy of $2.05$ MeV. This again is a result 
of the anharmonic nature of the collective potential. In table~\ref{tab:fe58gam}
we can see that the lowest state has a substantially reduced value of 
$\left<P_N\right>$ compared to the mean-field value at the minimum.
\begin{table}%[htb] 
\caption{The expectation values of the quadrupole- and pairing operators for
	$m=0$ collective Hamiltonian along the non-axial collective path in
	$^{58}$Fe and at the mean-field minimum.}
\label{tab:fe58gam}
\begin{ruledtabular}
\begin{tabular}{cdd}
 & Collective & Mean-field  \\
\hline
 $\left<Q_0\right>$ & 12.63 & 12.37  \\
 $\left<P_N\right>$ & 3.53 & 4.21  \\
 $\left<P_P\right>$ & 3.31 & 3.34  \\
\end{tabular}
\end{ruledtabular}
\end{table}

\subsection{Realistic application of large amplitude collective motion} 
The case of $^{58}$Fe has the advantage that the configuration space
is relatively small and therefor there are no big computational
problems, and we could compare exact and approximate solutions.  It
still allows us to explore several key features of our method and test
its feasibility and the quality of the results. To test the method in
more realistic circumstances we decided to apply our method to the
rare-earth region.  We have chosen $^{156}$Gd and $^{182}$Os since the
gadolinium nucleus is known to be $\beta$-soft, whereas the osmium isotope
is $\gamma$-soft. Both nuclei are situated in a region which is rich
in nuclear structure phenomena.

The calculations for $^{156}$Gd and $^{182}$Os are done in a configuration 
space consisting
of the $N=5(4)$ and $6(5)$ neutron(proton) shells. The spherical single particle
energies and suppression factors of Eq.~(\ref{eq:Qfactor1}) 
and~(\ref{eq:Qfactor2}) are again taken from~\cite{BK68}. 
We compare the RPA energies and the RPA vectors calculated with the 
full RPA and using the projected approximation in table~\ref{tab:bgpvib3}.  
\begin{table}%[hbt]
\caption{Comparison of the full RPA energy, $\hbar \Omega$, the projected
RPA energy, $\hbar \bar{\Omega}$, and $\delta_{B,1}$ for the
$\beta$-, $\gamma$-, $\Delta^{(1)}$- and $\Delta^{(2)}$-vibrations
in case of $^{156}$Gd and $^{182}$Os using the new projection basis  (\ref{eq:basis3}).  
The energies are in units of MeV. }
\label{tab:bgpvib3}
\begin{ruledtabular}
\renewcommand{\arraystretch}{1.2}
\begin{tabular}{cdddd}
 & \multicolumn{4}{c}{$^{156}$Gd} \\
\cline{2-5}
 & \multicolumn{1}{c}{$\hbar \Omega$} & \multicolumn{1}{c}{$\hbar \bar{\Omega}$} &
 \multicolumn{1}{c}{$\delta_B$} & \multicolumn{1}{c}{$\delta_1$} \\
\cline{2-5}
 $\beta$-vibration & 0.8850 & 0.9224 & 0.0040 & 0.5138\\
 $\gamma$-vibration & 1.6860 & 1.8490 & 0.0366 & 0.0325\\
 $\Delta^{(1)}$-vibration & 1.8089 & 1.8147 & 0.0049 & 0.0042\\
 $\Delta^{(2)}$-vibration & 1.9923 & 2.2256 & 0.9409 & 0.9339\\
\cline{2-5}
 & \multicolumn{4}{c}{$^{182}$Os} \\
\cline{2-5}
 & \multicolumn{1}{c}{$\hbar \Omega$} & \multicolumn{1}{c}{$\hbar \bar{\Omega}$} &
 \multicolumn{1}{c}{$\delta_B$} & \multicolumn{1}{c}{$\delta_1$} \\
\cline{2-5}
 $\beta$-vibration & 1.5704 & 1.5851 & 0.0058 & 0.0087\\
 $\gamma$-vibration & 1.1208 & 1.1458 & 0.0016 & 0.0010\\
 $\Delta^{(1)}$-vibration & 1.6690 & 1.6776 & 0.0106 & 0.0147\\
 $\Delta^{(2)}$-vibration & 1.8476 & 2.0673 & 0.2165 & 0.2826\\
\end{tabular}
\end{ruledtabular}
\end{table}
We find a good agreement for the low-lying solutions. The second
pairing vibration is somewhat too high in energy which is also reflected in a 
small overlap of the RPA vectors. The projection basis seems 
to work very well in the cases of heavier nuclei and larger configuration 
spaces examined here.

\begin{figure}%[htbp]
  \centerline{\includegraphics[clip,width=12cm,angle=0]{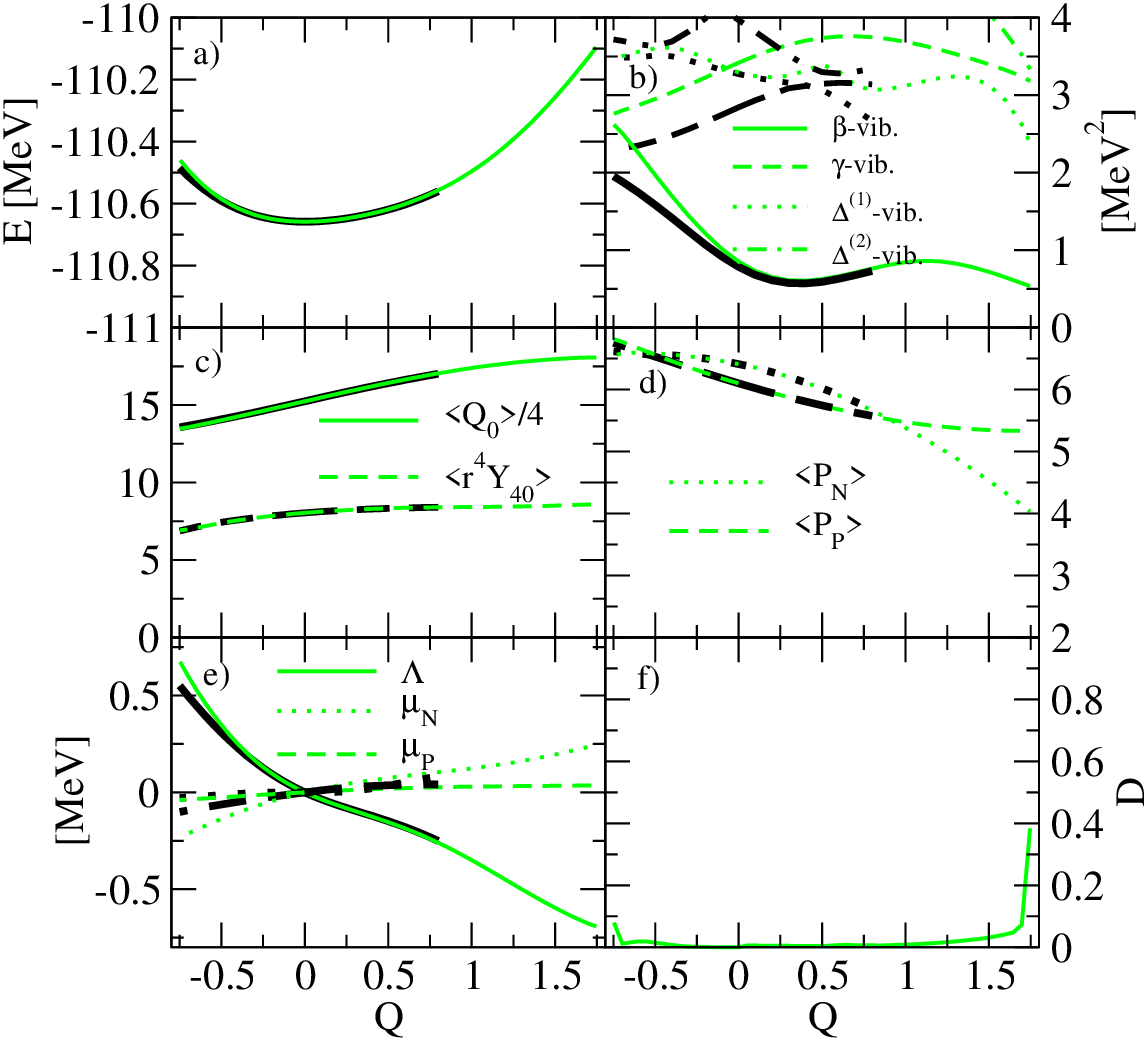}} 
  \caption{(Color online) Axial collective motion in $^{156}$Gd. See Fig.~\ref{fig:fe58beta} 
	for more details.} 
  \label{fig:gd156beta}
\end{figure}

\begin{figure}%[bhtp]
  \centerline{\includegraphics[clip,width=6.0cm]{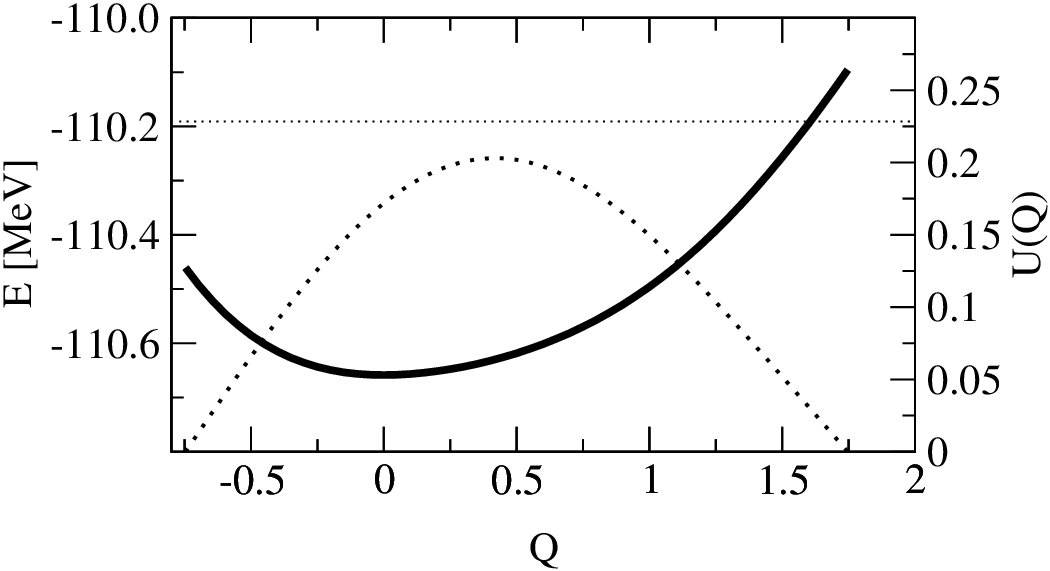}} 
  \caption{The energy along the collective path for projected axial large 
	amplitude collective motion in $^{156}$Gd (following the 
	first RPA solution) is drawn with the thick solid line with the 
	energy scale on the left side. The lowest eigenvalue of the Hamiltonian 
	is drawn as a thin horizontal line. The corresponding eigenfunction has 
	the scale on the right side.} 
  \label{fig:SEgd156}
\end{figure}

\begin{figure}%[htbp]
  \centerline{\includegraphics[clip,width=12cm,angle=0]{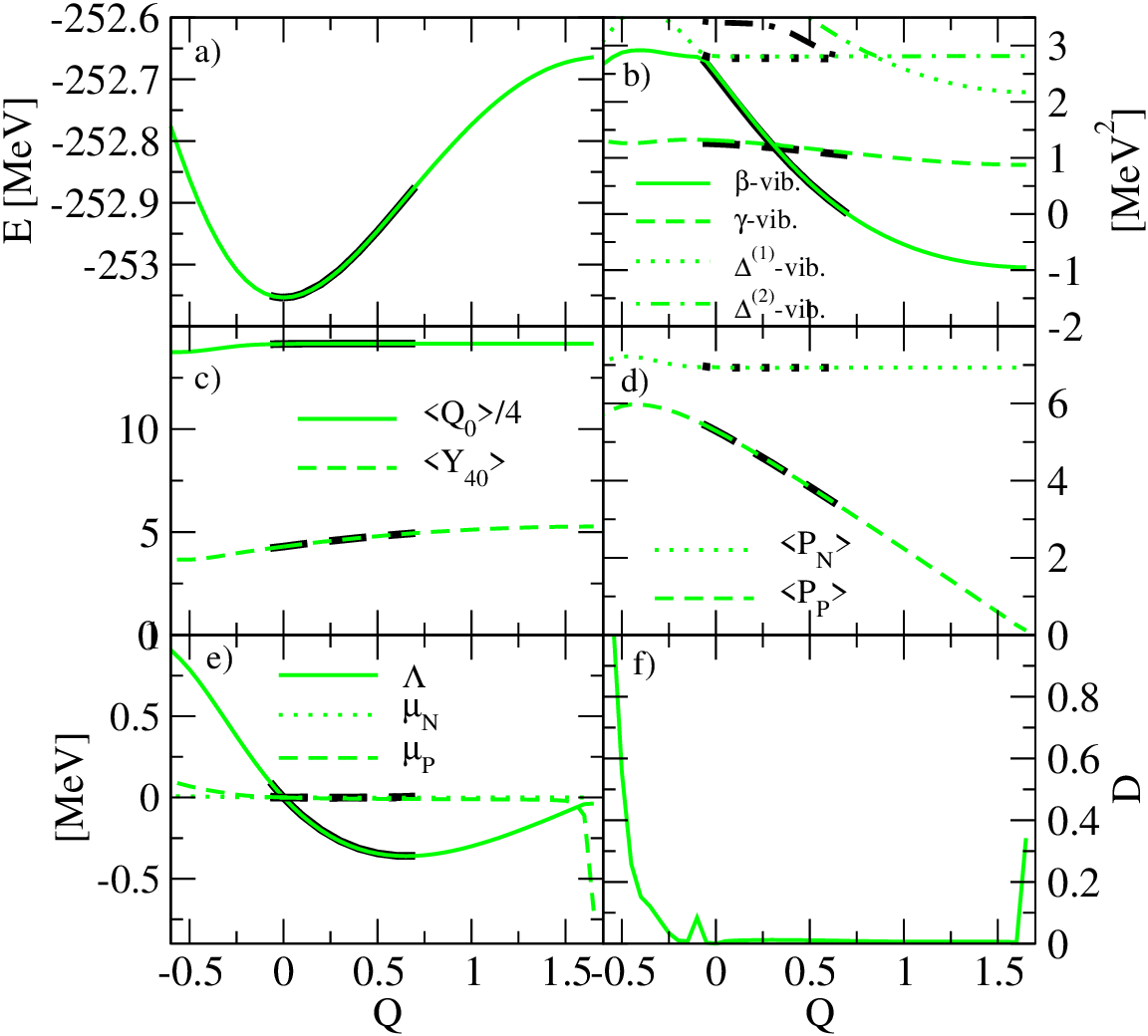}} 
  \caption{(Color online) Axial collective motion in $^{182}$Os. See Fig.~\ref{fig:fe58beta} 
	for more details.} 
  \label{fig:os182beta}
\end{figure}

In Fig.~\ref{fig:gd156beta}--\ref{fig:os182beta} we can see the results of 
the large amplitude collective motion following the lowest axial symmetric 
solution. We have included both the results obtained with the full RPA
as well those employing the RPA projected on a basis.
In $^{156}$Gd the lowest solution is the $\beta$ vibration but it also
has quite large pairing components, as can be  seen in
Fig.~\ref{fig:gd156beta} in the change in strength of the pair fields.
There is in general a non-trivial structure of the collective path close 
to the mean-field minimum which also can be seen in Fig.~\ref{fig:gd156beta} a)
where the energy show anharmonic behaviour around the minimum. 
Figure~\ref{fig:gd156beta} f) show a good decoupling of the collective degrees 
of freedom from the non-collective degrees of freedom in the region close to 
the minimum. 
In Fig.~\ref{fig:SEgd156} we can see that there is one low-lying solution 
of the collective Hamiltonian. 
The fact the lowest eigenvalue is situated high in energy, relative to the 
range in which we have found the collective potential, tells us that the 
assumption that $U(Q)=0$ at the at the ends of the collective path is not 
justified in this case. One can expect that the correct $U$ would stretch 
substantially outside the range on which we have calculated the collective path.

$^{182}$Os is a $\gamma$-soft nucleus and we show the result following the 2 
lowest normal modes.
In Fig.~\ref{fig:os182beta} we can see that the lowest axial RPA
solution is mainly of proton pairing nature. The strength of the
proton pair field is proportional to the collective coordinate and
that the pair field collapses at $Q\approx 1.6$ which  leads to a
jump in the chemical potential. For small negative values of $Q$ there
is an avoided crossing with a mode that is dominantly a shape
vibration, which leads to a reduction of $\left< Q_0 \right>$.
The energy along the collective path in Fig.~\ref{fig:os182beta} a) show a 
maximum when $\left< P_P \right> \rightarrow 0$ and a approximately harmonic
behaviour around the minimum.
Figure~\ref{fig:os182beta} f) shows a good decoupling of the collective degrees 
of freedom from the non-collective degrees of freedom in the region close to 
the minimum.
At large negative values of $Q$ we have a crossing with a proton paring vibration
which gives  large state mixing and therefore no decoupling of the 
collective solution.

\begin{figure}%[htbp]
  \centerline{\includegraphics[clip,width=12cm,angle=0]{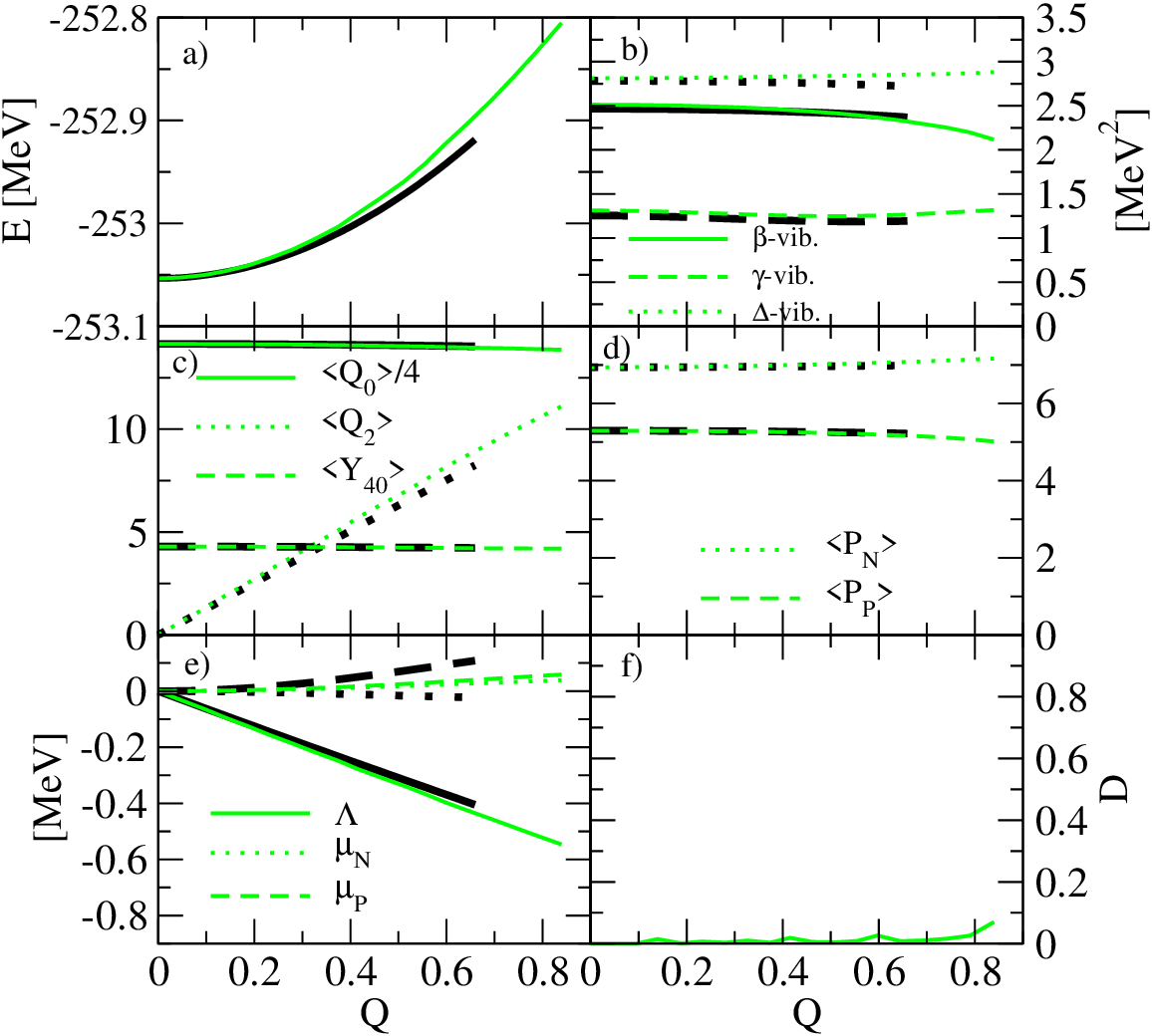}} 
  \caption{(Color online) Non-axial collective motion in $^{182}$Os. See Fig.~\ref{fig:fe58gam} 
	for more details.} 
  \label{fig:os182gam}
\end{figure}

Fig.~\ref{fig:os182gam} show the results when following the 
collective path defined by the lowest $\gamma$-vibration in $^{182}$Os.
The calculation shows that the collective path is mainly dominated by the increase 
of the $\left< Q_2 \right>$ tri-axial deformation. At $Q> 0.8$ we see an 
avoided crossing of the $\beta$- and $\gamma$-vibration which causes an numerical 
instability in our calculations. This also signals the need for more then one 
collective coordinate. Even though there are numerical difficulties in 
implementing our method in some cases 
we can see that our projection method works very well in the larger 
configuration spaces employed here and it is practically implementable.

\section{Conclusions and summary\label{sec:Conclusions}}
We have extended the method of calculating the self-consistent collective 
path presented 
in~\cite{DK00} to include constraints on the particle number and implemented 
it for the quadrupole+pairing Hamiltonian~\cite{BK68}. The method consists of 
finding a series of points fulfilling the force equation, where the local 
direction of the collective path is determined in each point by the local
normal modes. The local RPA equations and the force equation are solved in a 
double iterative process with constraints on the particle numbers and the step
length along the collective path. The method allows us to determine the collective 
coordinate from the Hamiltonian without having to assume a priori which are 
the relevant degrees of freedom.

To implement this method in heavier systems and for more 
realistic nuclear forces we need to truncate the RPA calculation in a way that 
will give an accurate approximation of the low-lying RPA solutions. 
We have improved the projection method originally presented in~\cite{NW99}
in such a way that we are able to describe all low-lying states including 
pairing vibrations. This is done by expanding the state-dependent basis 
suggested in~\cite{NW99} to include a pairing term, which is only active around 
the Fermi surface. The new basis set gives RPA energies very close to the exact 
solutions. The calculation of the overlap of the wave functions between the 
full and our approximate RPA solution shows that the wave functions are almost 
identical. We can therefore expect our method to give a good 
approximation to the collective path.

Our method of calculating the collective path has been implemented for 
the cases of $^{58}$Fe, $^{156}$Gd and $^{182}$Os. 
We have chosen to follow the lowest axially symmetric and tri-axial solutions. 
The decoupling of our collective coordinate from all the non-collective 
coordinates can be quantified in the decoupling measure $D$. This is found to be 
small along the collective path with the exception of regions of avoided crossings
where the system undergoes configuration mixing. In such regions one collective 
coordinate is not sufficient to describe the system accurate. 
In regions where the projection basis gets over-complete $D$ has to be 
calculated with special care. This happens when the proton and/or the neutron 
pairing collapses.

We see that the collective path goes through avoided crossings with pairing 
solutions in most cases in both $^{58}$Fe and $^{182}$Os. This leads to collapses 
of the pair fields and an end of the collective path. These avoided crossings also 
show that more then one collective coordinate would be needed for a accurate description 
of the collective motion. In the $\beta$-soft nucleus $^{156}$Gd the collective 
coordinate is of a more complicated structure which can also be seen in the 
non-harmonic shape of the potential energy.
Our projected local RPA method for calculating the collective path gives very 
good agreement with the results obtained using the full RPA. The method is 
very useful when calculating self-consistent large amplitude collective motion
in large a configuration space.

By solving the one-dimensional ``radial'' Schr\"{o}dinger equation
along the collective path we are able to examine the effect of the
collective motion on ground state properties. In cases where the
collective path ends with a collapse of the pair field we must include
the effect of pairing rotations on the low energy spectrum. It is
surprising that almost all our calculations are dominated by states with
collapsing pairing; there may well be important lessons in this
feature.  The reason they occur so frequently is the presence of
low-lying configurations without pairing. One might ask whether this
is an artifact of our model, and whether larger configuration-spaces
with more complicated interactions would behave differently. Such
calculations are clearly called for, but we do not expect dramatically
different results, since pairing mainly acts in a small region around
the Fermi-surface. There is also a slight possibility that the approximations
we made in our treatment of spurious admixtures contributes to these 
effects. The surprising importance of the pairing collapse
is the main result of our calculations that is usually not seen in a
standard constrained mean-field calculation.

In this paper we have implemented a method to find the adiabatic
self-consistent collective path for a nuclei. A technique to truncate
the basis in which the RPA equations are solved has been improved and
a good agreement between the full and truncated RPA is found. To solve
the RPA equations in a limited basis has proven to be a useful and
practical way of calculating the collective path within the local
harmonic approximation. We intend to apply similar techniques to the
interesting problem of collective motion at finite rotational
frequency in the near future. It remains to be investigated how we can
include the covariant terms in the RPA in a suitable approximation.

\section*{Acknowledgement}
This work was supported by the UK Engineering and Physical Sciences Research 
Council (EPSRC) under grant GR/N15672.

\section*{Appendix}
The six conditions (Eqs.~(\ref{eq:dQ1}) and~(\ref{eq:dRN1}--\ref{eq:dRN2})) 
gives a set of equations which can be solved for the 
six parameters $\epsilon_i^{(n)}$, $\eta_{i \tau}^{(n)}$, $\lambda_i^{(n)}$ 
and $\mu_{i \tau}^{(n)}$ for each $i$ and $n$. For $i=0$ we get
\begin{eqnarray}
	\label{eq:eps0}
	\epsilon_0^{(n)} &=& \frac{\Delta Q}{1 + \frac{f^{(0)} \cdot 
	f^{(n)}}{f^{(n)} \cdot f^{(n)}} - \sum_{\tau} \frac{N^{20(n)}_{i \tau} 
	\cdot f^{(n)}}{f^{(n)} \cdot f^{(n)} N^{20(n)}_{i \tau} \cdot 
	N^{20(n)}_{i \tau}} \left( f^{(0)} + f^{(n)} \right) \cdot 
	N^{20(n)}_{i \tau}} , \\
	\label{eq:eta0}
	\eta_{0 \tau}^{(n)} &=& - \epsilon^{(n)}_0 \frac{f^{(n)} \cdot 
	N^{20(n)}_{i \tau}}{ f^{(n)} \cdot f^{(n)}} .
\end{eqnarray}
For all other $i$ 
Eqs.~(\ref{eq:dRni1}) and~(\ref{eq:dRni2}) together with the 
constraints~(\ref{eq:dQ1}--\ref{eq:dRN2}) give
\begin{eqnarray}
	\label{eq:epsni}
	\epsilon_i^{(n)} &=& \frac{-\delta^{(n)}_i f^{(0)} \cdot 
	\Delta^{(n)}_{\perp i} {\cal R} + \sum_\tau \frac{N^{20(n)}_{i \tau} \cdot 
	\Delta^{(n)}_{i-1} {\cal R}}{N^{20(n)}_{i \tau} \cdot N^{20(n)}_{i \tau}} 
	\left( f^{(0)} + f^{(n)} \right) \cdot N^{20(n)}_{i \tau}}{1 + 
	\frac{f^{(0)} 
	\cdot f^{(n)}}{f^{(n)} \cdot f^{(n)}} - \sum_\tau \frac{N^{20(n)}_{i \tau} 
	\cdot f^{(n)}}{f^{(n)} \cdot f^{(n)} N^{20(n)}_{i \tau} \cdot 
	N^{20(n)}_{i \tau}} \left( f^{(0)} + f^{(n)} \right) \cdot 
	N^{20(n)}_{i \tau}} , \\
	\label{eq:etani}
	\eta_{i \tau}^{(n)} &=& - \epsilon^{(n)}_i \frac{f^{(n)} \cdot 
	N^{20(n)}_{i \tau}}{f^{(n)} \cdot f^{(n)} } - N^{20(n)}_{i \tau} \cdot 
	\Delta ^{(n)}_{i-1} {\cal R} , \\
	\label{eq:lambdani}
	\lambda^{(n)}_i &=& \frac{H^{20(n)}_i \cdot f^{(n)} - \sum_\tau \frac{ 
	H^{20(n)}_i \cdot  N^{20(n)}_{i \tau}  N^{20(n)}_{i \tau} \cdot f^{(n)}}{
	N^{20(n)}_{i \tau} \cdot N^{20(n)}_{i \tau}}}{f^{(n)} \cdot f^{(n)} - 
	\sum_\tau \frac{\left( N^{20(n)}_{i \tau} \cdot f^{(n)}\right)^2 }{
	N^{20(n)}_{i \tau} \cdot N^{20(n)}_{i \tau}}} , \\
	\label{eq:muni}
	\mu^{(n)}_{i \tau} &=&  \frac{H^{20(n)}_i \cdot N^{20(n)}_{i \tau} -  
	\lambda^{(n)}_i f^{(n)} \cdot
	N^{20(n)}_{i \tau}}{N^{20(n)}_{i \tau} \cdot N^{20(n)}_{i \tau}} .
\end{eqnarray}
These equations can easily be generalised to incorporate additional
constraint operators like angular momentum. They are slightly more
complicated than those shown in other work~\cite{DK00}, since we have
chosen to fix the step size in the collective coordinate.

\end{document}